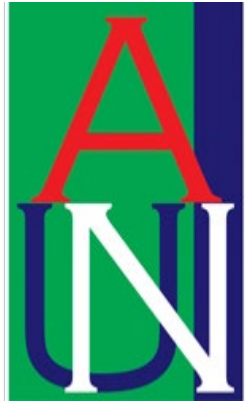

# INVESTIGATION OF HOW SOCIAL MEDIA INFLUENCED THE ENDSARS PROTEST IN LAGOS, NIGERIA

CHRISTOPHER AUGUSTINE

A00022327

Submitted in accordance with the requirements for the degree of

Master of science in information systems

American University of Nigeria, Yola

Department of Information Systems

School of Information Technology & Computing

MAY, 2022

## Intellectual Property and Publication Statements

I, Christopher Augustine, confirm that the work titled "Investigation of how Social Media Influenced the EndSARS Protest in Lagos, Nigeria" is my work and that appropriate credit has been given where reference has been made to the work of others.

This copy has been supplied on the understanding that it is copyright material and that no quotation from the thesis may be published without proper acknowledgement

# Statement of Committee and Institutional Approval

We, the undersigned confirm that this research was conducted and written by
(Christopher Augustine) under our supervision.

(Signature and Date)

**Dr. Samuel Utulu**

**Thesis Supervisor**

School of Information Technology & Computing

(Signature and Date)

**Dr. Narasimha Rao Vajjhala**

**Chair of Program**

School of Information Technology & Computing

(Signature and Date)

**Dr. Rabiu Ibrahim**

**Graduate Program Coordinator**

School of Arts and Science

(Signature and Date)

**Dr. Abel Ajibesin**

**Dean of Host School**

School of Arts and Science

# Acknowledgments

This thesis is dedicated to Almighty God, My parent Engr, and Mrs. A.C Okoye. Not forgetting my lovely brothers

I want to express my genuine gratitude to Almighty God for his mercy and help throughout my years of academic challenge; my profound gratitude also goes to my beautiful parents Engr. & Mrs. A.C Okoye, my supportive brothers Valentine, Stanley, Henry, Cyprian, and Michael, I appreciate the encouragement and prayers.

Special gratitude and respect to my amiable supervisor and mentor, Dr. Samuel Utulu, Associate Professor of Information systems, who molded me academically via constant moral, fatherly support and encouragement during the rigorous process of writing my master's thesis. Prof, I enjoy your continuous mentorship; all graduate students love you.

I am especially indebted to departmental Professors, who have supported my career goals, and worked actively to provide me with the protected academic time to pursue those goals. I am grateful to all of those with whom I have had the pleasure to work during this and other related projects.

# Abstract

Social media are tools for digital communication and information exchange. Social media have been used to promote awareness of various socio-political, economic, and environmental issues efficiently and effectively. Nigerians use social media platforms, including Facebook, Twitter, WhatsApp, Instagram, etc., to interact with one another and shape public discourse and policies. Hence, social media has become means for communication and raising awareness among citizens and government about public policies. A good example is a petition on social media on disbandment (End) of the Special Anti-Robbery Squad (SARS) of the Nigerian Police Force. Consequently, the research investigates how social media impacted the EndSARS protests in Lagos, Nigeria. The study was necessary because most of the reflections available in the literature on social media-driven protests are based mainly on dualist assumptions. This is to say that scholars present technology (social media), everyday life experiences that lead to protests, and the societies (contexts) where the protest occur as different entities. This is not in line with insights in the socio-materiality theory that is driven by notions in the interpretivism philosophy. Consequently, this study aims to carry out a survey that will show how social media, everyday life experiences, and contexts where protests were held are interwoven. The aim will enable the presentation of a more appropriate explanation of phenomena surrounding the impact of social media on protests. The study is an inductive qualitative study and adopted the interpretive case study. The research context was selected based on the convenient sampling technique, while the study participants were selected based on the snowball sampling technique. Data were collected from study participating using the unstructured in-depth interview technique. Data collected in the study were analyzed using the thematic data analysis technique. Study findings reveal how the Nigerian socio-political context-SARS brutality, unemployment, corruption, bad governance, insecurity, and misuse of public power played out in real-life contexts and on social media and how the embeddedness of events in real-life contexts and the social media impacted the EndSARS protests in Lagos, Nigeria. I conclude that the unfriendly Nigerian socio-political caused the EndSARS protests. Social media only provided an extended platform for experiencing the Nigerian socio-political environment and the protests.

# Table of Contents

# List of Tables

# List of Figures

# Chapter 1

# Introduction

## 1.1 Background of the study

The use of social media as a medium of communication has become popular in political, economic, social, and commercial circles, including in Nigeria. Individuals have become interested in social media as a method of self-expression and communication. Social media platforms such as Facebook, Twitter, WhatsApp, Instagram, among others, have been widely used in Nigeria as media for communication and sharing information in various formats. The habit of sharing personal and group pictures and videos, textual and oral messages has become entrenched in people's daily lives (Allcott & Gentzkow, 2017). People use varying formats of communication made available on social media to show their hobbies, enjoy entertainment, and tell personal stories about their experiences and life. The use of social media has aided the transmission of information by individuals and business organizations, social movements, and organizations involved in social marketing and public policy communication. Good examples include the bring-back-our-girls group, black-life-matters group, and similar groups established to agitate social change (Ghonim, 2012).

In the recent past, the EndSARS group was also established by concerned Nigerian youths and made its modes of operations social media-based (Salaudeen, 2017). Nigerian youths set the EndSARS group to communicate to the public and government about police brutality and the need for comprehensive policy and police reforms in the country (Salaudeen, 2017). This followed the Black-Life-Matters Group in the US, which was established to agitate for policy reforms due to increased police brutality in the US (Garza, 2014). Similar protests like those in Singapore, Myanmar, Iran, Hong Kong, and Thailand (Barbera, 2015). Like the groups preceding it, the EndSARS Group used social media to publicize its anti-police brutality agitations. A new trend manifests in how people use social media to air their views on public issues and opinions, recruit public support, and promote protests.  The recent trend necessitates a thorough scholarly examination of how social media impacts street protests in Nigeria and elsewhere around the globe.

The EndSARS protests held in Lagos and other parts of Nigeria brought to light the social, economic, and political injustices youths suffer. On 3 October 2020, online agitation against the Nigerian Police Force's Special Anti-Robbery Squad (SARS) squad was revived when a video showed several SARS operatives leaving a scene in a white car that belonged to an unarmed young man in Delta state. It was reported that the SARS operatives shot the young man (Amaza, 2020). Youths saw the incident as an opportunity to revisit their campaign, which started in late 2017 against SARS brutality. They were to use the option to inform the global community about injustices they suffer due to the bad governance and lack of respect for human rights that resulted in SARS brutality and the government's inaction towards putting an end to SARS brutality. The shutdown by the Central Bank of Nigeria digital currency donation sites used by those championing the anti-SARS brutality protests further showed how far the Federal Government of Nigeria is willing to stop SARS brutality. It also showed how far the Federal Government of Nigeria was ready to go to help Nigerian youths get the justice they deserve against SARS operatives. However, efforts made by Nigerian youths proved more decisive than past efforts, given the adoption of social media as grounds for agitation and protests (Harper, 2020).

Issues regarding military intervention against protesters at the Lekki toll gates, curfew order issued by the Lagos State Government, and firing of teargas by men and officers of the Nigerian police force, which caused stampedes and, deaths all showed how to open Nigerian governments, were too democratic principles of defending human rights to protests and agitations. This occurred despite the declaration by former Inspector-General Ibrahim Idris that the SARS should not employ torture or extrajudicial actions against protesters (Punch, 2013). Amnesty International also provided credible revelations that showed that SARS operatives tortured and abused protesters despite the police Inspector General's directives (Amnesty International, 2014). Amnesty International's revelation shows that indicates rights abuses against protesters have been an age-long issue in Nigeria persist (Amnesty International, 2014). The EndSARS protests show that problems relating to bad governance, SARS brutality, misuse of power, unemployment, corruption, and insecurity, which are enshrined in the Nigerian socio-political environment, persist despite the socio-technological advancements that led to the invention of social media. However, the announcement of "urgent restructuring" of SARS operations and an official inquiry into all allegations levied against the unit resulted from pressures from the international community and showed how efficacious social media was to achieve the organizers of the EndSARS protests. The implication is that three factors central to the social materiality theory propounded by Orlikowski (2007) come to bear in the impact of social media on the EndSARS protest in Lagos, Nigeria. Orlikowski (2007) argues that information technology, everyday life experiences, and contexts are embedded and should be given credence when studying how information technology impacts human actions.

This research seeks to assess how social media impacted the EndSARS protests in Lagos, Nigeria. The study implies that it will produce theoretical insights useful for both theory and practice. This is to say that scholars will find them valuable studies for further research. Policymakers and those engaged in cyber-activism will also find the practical study as insights derived from the study will provide a research model that closely represents social realities surrounding the impact of social media on protests.

## 1.2  Problem Statement

It follows that history demonstrates that protests against government policies have always occurred in Nigeria. The Aba Women Protest, SAP Riot, and Ali Must Go protests are good examples. Consequently, in Nigeria, protests predate the invention of information technology and social media. The EndSARS protests in Nigeria in 2020 were another striking illustration of how individuals may come together to protest against bad governance, social injustice, insecurity. The difference between the EndSARS protests and those that preceded them is that Nigerians, especially young people, utilized    media to mobilize for the replacement by the government of SARS. Social media were used to spread awareness about the EndSARS campaign. Across the globe, the use of social media as a platform for agitating against governments has become popular. This gave rise to scholarly works on how social media impact protests organized against governments and other related themes (Zeynep, 2014). The implication of this is the evolution of new forms of scholarly done to expose and understand the impact of social media on activism, protests, and riots. Interestingly, much has been done regarding theoretical and practical needs to uncover and understand social media's impact on activism, protests, and riots.

The question that comes to mind is how appropriately have existing studies done regarding adequately exposing and understanding how social media has impacted protests against governments in the recent past. The concern and problem are very relevant to the information systems discipline. Research problems usually revolve around theoretical and practical issues in the information systems discussed. In most cases, theoretical issues are conceptualized to include methodological and philosophical issues. Consequently, this study addresses both theoretical and practical problems. In other words, the problem central to the study is the pressing need to expose and understand how social media are embedded into social contexts and how this influences how social media impacts protests against governments. Most studies that were carried out on how social media impact protests are based on dualist perspectives (Bardici, 2012). In other words, the studies conceptualize social media and social contexts where protests take place as dual entities, that is, as different entities. This is to say that first, the everyday life realities on social media, and second, the everyday life realities in real-life physical contexts. The underlying philosophy adopted in the studies promotes the inclination to conceptualize social media and social contexts where protests take place as dual entities.  Most of the studies are positivism inclined and therefore conceptualized social media and social contexts as dual and objective entities (Petter et al., 2007). For instance, several studies have shown that information communicated through social media may turn people against their government (Conor, 2019; Eze, 2020). The studies seem to argue that what happens on social media is only determined by the everyday life realities experienced on social media by social media users. This is problematic because everyday life realities are embedded on social media and everyday realities in real-time in physical contexts.

Another problem that informed the study is the dominance of deductive reasoning in studies on how social media impacts protests against governments. While one may not claim that using deductive reasoning is problematic, one may argue that an adequate context-based understanding of how social media impacts protests against government is likely to be promoted by studies that adopt inductive reasoning. This is because inductive reasoning-based studies collect data without formal theories (Heit, 2000). Using formal theories allows using a posteriori knowledge based on previous experience (Honderich, 1995). Undoubtedly, adopting existing formal theories (a posteriori knowledge) introduces some bias that does research titled towards the assumptions in the theories. The implication is that deductive reasoning-based studies are likely to replicate existing inherent beliefs in the formal theories they adopt. This makes it very difficult for research done in contexts alien to the formal theories adopted in deductive reasoning-based study to develop novel and context-specific findings. In the information systems discipline, the call for scholars to build context-specific theories, otherwise known as middle-range theories, has been persisting (Bhattacherjee, 2012). At the center of the call reasoning for the mid-range, theories need to develop relevantly, and this call may be quickly reached if scholars adopt inductively. Issues such as this warrant the reconsideration of the research approach adopted for studies that are carried out to understand the impact of social media on protests against governments. This is particularly important if the protests occur in developing countries, such as Nigeria, where democratic principles are interpreted and implemented differently from those of more matured democracies in Europe and the US.

## 1.3 Research Objectives

The study's primary objective is to address the problems discussed in the segment above, namely, theoretical and practical issues. More specifically, the student contexts seek to address the dualist approach to assessing the impact of social media on protests against governments. The study's objective is to explain the embeddedness of social media, the everyday life experiences of protesters, and the context of the protests. The study uses the EndSARS protests in Lagos, Nigeria, as a case study. Consequently, the study's objective is to explain the embeddedness of social media, the everyday life experiences of protesters, and the Nigerian socio-political context as experienced in Lagos, Nigeria. This will provide grounds to conceptualize social media as socially constructed information technology and explain how social media impacted the EndSARS protests.

## 1.4 Research Questions

The following broad research question was coined to achieve this objective: how did social media impact the EndSARS protests in Lagos, Nigeria? Two specific research questions were coined to address the general research question. These include:

1.  How did the embeddedness of socio-political contexts in Nigeria and the social media impact the EndSARS protests in Nigeria?
2.  How did the embeddedness of the everyday life experiences of protesters and the social media impact the EndSARS protests in Nigeria?

## 1.5 Significance, Scope, and Limitations of Study

Research studies like this one are helpful firstly to the academic community. The study makes a theoretical contribution to the information systems discipline. Specifically, the study makes a theoretical contribution to information systems in developing countries. The study serves as an addition to existing studies and theoretical insights on the impact of information systems on protests. Also, to create an environment that promotes and protects the rights of citizens, the rule of law, protection of lives and property, as stated as the primary function of security personnel, this study will assist the government and policymakers in aligning and realigning laws and policies to meet the needs of the populace. The study will also assist in limiting the abuse of power by security personnel in the country. The subject scope of the study is to investigate the impact of social media on the EndSARS protests in Nigeria in 2020. Residents in Lagos State, Nigeria, will focus on the investigation. The participants and location selected are activists and protesters in Lagos, Nigeria, where the protest occurred. Lagos is a vast state so, the study parts where the protest occurred will center our study. The limitation of the study was that the focus was mainly on protesters and activists in Lagos, Nigeria. The study left out other cities in southern Nigeria that actively participated in the EndSARS protests. Consequently, the study may not lay claim to generalizable sample size. However, the study makes theoretical generalizations.

# Chapter 2

# Literature Review

## 2.1 Introduction

In this study segment, I present the outcome of a preliminary assessment of relevant and related literature available in the literature. The piece resulted from the literature review to produce and deliver the study proposal. This is given that the study adopted the inductive reasoning approach where a comprehensive review of the literature is not applicable. The segment begins with an assessment of the history of protests in Nigeria, dating back from the colonial era to the post-colonial period. It also looked into social media, protests, and the EndSARS.

## 2.2 History of Protest in Nigeria

Like most nations under European domination, Nigeria has suffered violent riots and uprisings throughout its history, from colonial to post-independence. There has been a long history of militant warfare in Nigeria and European occupation and subjection, making it difficult for certain authorities, such as the Sokoto Caliphate, the Ijebu, and the Benin Kingdom, to be captured and surrendered.

Following the fall of indigenous rulers, the colonial state instituted a system of indirect control that supported European interests by guaranteeing social order via administration by provincial authority. The Aba Women's Revolt of October 1929 began a sustained campaign against this system. Around 10,000 women demonstrated against a proposal by British authorities to tax women directly, apart from males. The ladies launched attacks against native courts, warrant officers, and industries across Europe. The British colonial government chose to put down the insurrection by force, which resulted in the deaths of hundreds of demonstrators and the destruction of their property. Even though the insurrection was put down, the imperial government decided to scrap plans for the Warrant Chief System, taxing women (van Allen 1972). People engaged in economic and political protest during the Aba Women's Revolt, which resembled other forms of civil disobedience in Nigeria throughout the twentieth century.

The Niger Delta was historically one of nine critical centers of Acritical can colonial resistance. In addition to crop delays, primary measures pursued included tax evasion and boycotts, industrial action, worker demonstrations, autonomous African churches, and cultural/welfare groups, shattering the monopoly of European corporations, and occasional revolts and riots. Women's social protests were unusual, although they did occur (Alder, 1999; Mba, 1982). The projected taxation of women triggered the Aba riots or women's war of 1929 of British Colonialism's Indirect Rule agenda in Nigeria (Afigbo, 1966; Journal of African Society, 1930, pp. 542–543). This was preceded by extensive wealth assessments, population, livestock, and economical tree counts. For a long time, taxes were blamed for the unrest. Others included: persecution, extortion, and corruption among Native Court members, high-handedness of appointed warrant chiefs, illegal and oppressive sanitary fines, continued and enforced unpaid labor on civil constructions, unfair/excessive imprisonment, low prices of farm products (especially palm oil and palm kernel) versus high prices of imported goods (Afigbo, 1972; Arifalo; Davies, 1960;). Men in many sections of southern Nigeria had been taxed repressively a year earlier, in 1928. The colonial government was shaken by the women's protest, forcing a review of the law. According to historian Obaro Ikime, the Aba demonstrations were "protests against the total of complaints linked with present British administrative procedures and the concomitant advances of western civilizations" (Ikime 1980, p. 444). The protests started on November 18th and lasted over three months. They included four divisions in Owerri, two divisions in Calabar, and the Afikpo division in Ogoja. Protests were especially fierce, with significant commerce and market routes frequented by female merchants. The contact between women from diverse origins was fostered by long-distance trade, Ade-Ajayi & Espie (1965, pp. 203, 394). Most traditional African communities are significant social, political, and cultural hubs. From 1941 through 1947, Egba matriarchs committed civil disobedience, rallies, and insurrections against colonial exploitation, taxes, market closures, and product delays. So the colonial authorities and their local agents were kidnapped, and the traditional monarch, Alake, was dethroned.

Nigeria's post-independence era began with hostility between its three regions, with the southern West and East regions fearing and resisting the powerful Hausa-Fulani North. Until 1966, when over 30,000 southerners (mainly Ibos) were killed in Kano, a major metropolis in northern Nigeria, the political landscape remained tense. Fearing for their safety, the Ibos moved in masse to the east. The Ibos' resolve to split and create the independent state of Biafra led to a civil war from 1967 to 1970, with an estimated 2 million deaths. Nigerian federal troops suppressed the secessionist movement, while General Gowon, the country's military leader, offered a program of reintegration, reconciliation, and rebuilding. This endeavor to reintegrate and restore in the Nigerian federation has failed. The country's ethnic/religious divisions and insurgencies have prevented national unification. An insurrection organized by Mallam Muhammed Marwa in December 1980 killed 4,177 citizens, 100 police officers, and 35 military officials (Aluko 1985). In March 1987, a clash between Christians and Muslims at the Kafanchan College of Education expanded to Zaria, Kaduna, Katsina, and Funtua, resulting in further fatalities (Ibrahim, 1989). Around 2,000 Christians tried to demonstrate against the legislation in Kaduna in February 2000 peacefully but were massacred in a horrific fight that triggered retaliatory murders throughout the nation. Other notable ethnic clashes include the Tiv-Junkun, Aguleri-Umuleri, and Ife-Modakeke disputes. Nonetheless, civil society upheavals occur outside ethnic/religious warfare in post-colonial Nigeria. People have protested against government policies. As a result of widespread arrests in September 1969, the Agbekoya Parapo (Farmers Resist Suffering Society) opposed implementing an £8 per farmer tax rate, among other trade concerns. Despite the deaths of government officials and farmers in the subsequent struggle, the farmers forced the government to compromise (Adeniran,1974). Similarly, the neoliberal International Monetary Fund Structural Adjustment Program introduced by the General Babangida Regime in 1986 led to significant student revolts in April, May, and May 1989. Initially, the government closed institutions and banned the National Association of Nigerian Students (NANS). The government ultimately adopted an SAP relief package, but insufficiently to reverse the escalating poverty and social misery among peasants and workers (Shettima, 1993). Devastating problems followed the suspension of the June 12, 1993, presidential election, which presumed winner Chief M. K. O. Abiola had won. General Babangida finally gave authority to a temporary administration headed by Chief Shonekan, but General Abacha overthrew it after around 90 days. When Abacha opted to stay in power, the opposition founded the National Democratic Coalition (NADECO) in May 1994, which coordinated demonstrations against his regime until his untimely death on June 8, 1998, clearing the door for a democratic transition. Since the 1990s, the Niger Delta area of Nigeria has become more aware of injustice and environmental damage caused by substantial international multinational businesses' oil extraction. The 1990 riots erupted in response to the Abacha regime's treatment of Ogoni leader Ken Saro-Wiwa, who was arrested and executed in November 1995. The Nigerian military government's death of Saro-Wiwa did not quell the Niger Delta uprising but instead fueled it. By the early twentieth century, the Niger Delta was teeming with militant organizations, some of which had taken up weapons against the Nigerian state. Nigerians' opposition to government and military brutality has grown in the early twenty-first century. The Peace and Development Project (2006) evaluated approximately 400 violent disputes in Nigeria between 2004 and 2005, mainly caused by injustice and inequality.

After General Babangida Junta annulled the June 12 presidential elections on June 26, 1993, the NLC's Central Working Committee convened in Lagos on June 28. It released a thorough assessment of the regime's justifications for annulling the polls. The NLC blamed the military government for derailing democracy in Nigeria, confusing the nation, tinkering with the transition program, and tolerating the excesses and failures of politicians, including the two presidential candidates during the campaigns, only to use the issues it had accepted (even encouraged) as excuses for precipitating a political stalemate. It warned that the escalating political turmoil would inevitably lead to significant economic problems "Unrealized hopes for bettering the lives of Nigerians would stay unmet. In this context, the worsening political crises must be connected to worsening poverty and threats to peace and order, which directly affect trade unions and the Nigerian working class "(Labour Congress 1993). Announcing its opposition to the military, the NLC indicated that it was ready to join a public battle for military withdrawal from politics and the restoration of democracy.

After its national executive council meeting in port Harcourt NLC then sent a list of requests to the Babangida junta in July 1993. A new presidential election was not "affordable" because the country could "not afford another presidential election with its attendant wastages, apathy, controversy, lack of faith and credibility," the Congress said, demanding the results of the June 12 election be released and supporting pro-democracy forces' arguments that the Nigerian crisis had nothing to do with region, religion, or tribe. "Return to complete and unconditional democratic administration by August 27, 1993," the NLC demanded. It warned the two government-created political parties "against being pressured into a hopeless election" and threatened to "see any such surrender by the parties as a betrayal of the Nigerian people." The Congress threatened to "question the legitimacy of the military" if it prolonged its stay beyond August 27, 1993, calling it "a recolonization of the Nigerian people...by the military" (Akin-Olukunle 1993). Finally, the communique called for the immediate release of detained activists and leaders of pro-democracy movements and proclaimed the "commitment of Nigerian trade unions to the unity and corporate.

On July 9, 1993, the NLC issued a strike warning giving the military junta a 12-day ultimatum demanding an end to political arrests and detentions and the release of SDP presidential candidate Chief Moshood K. O. Abiola. The military administration refused to satisfy these requests and instead invited the NLC to discuss them. The NLC threatened a nationwide strike if the troops did not leave by August 27 and demanded that "the Federal Government proclaim the 1989 constitution by August 27, 1993." (Komolafe 1993a; Vanguard 16 August 1993). To the government's communications secretary, Uche Chukwumerije's charges that unhappy politicians were inciting the Congress, Morgan Anigbo (The Guardian 25 August 1993) released a statement stating that " "congress' concerns transcend racial, tribal, and religious divisions. "It is regrettable that such mistaken accusations come from a patriot and nationalist." "

As a result of enormous political activity organized by the Campaign for Democracy, General Babangida resigned on August 27, 1993. (CD). He installed an Interim National Government (ING) with military commanders and no legal or constitutional basis. The ING, led by businessman Chief Ernest Shonekan, lacked legitimacy, was publicly rejected by SDP governors, was not recognized by the international community, and could not satisfy the NLC's demands.

In most Nigerian colleges, student conflict is the most intractable. Few universities ended their academic sessions without violent student protests in the previous decade. According to Alimba (2008), the first violent student protest occurred in 1971 at the University of Ibadan, Nigeria, killing Kunle Adepeju. Several hostile and destructive student protests followed the growth of higher education in Nigeria, making the educational environment unsuitable for teaching, research, and public service. The most concerning student discontent are its relentless nature and inherent violence. Student disputes outnumber other conflicts regarding frequency, volatility, and impact on institutions and the country (Aderinto1994). Consequently, barely any university administration has seen any type of confrontation, whether internal or foreign. According to Tayo (2006), the Nigerian university system has been plagued by student-related conflicts, alarming many stakeholders. This nasty trend is so common that many people accept it as part of a university education. Between 1989 and 1977, severe unrests and outbursts by Nigerian university students were reported to have doubled. Students' disagreements are becoming an endemic characteristic of the Nigerian educational system, according to Okoge (1992) and Ogunyemi (1994).

They noted the staggering loss of life and devastation of public property caused by student unrest. Thus, teenage violence on post-secondary campuses has become a concern for university administrators, governments, and members of civil society. Because tertiary institutions are subgroups of macro organizations, they reflect society's growing use of violent methods to settle disagreements. With over 150 higher institutions, Nigeria has seen unusual aggressive behavior by students involved in all types of disputes and violence.

Mohammed (2005) found that between 1986 and 1996, around thirty-three students and seven academic staff members were slain in student-on-student violence. According to Rinju (2003), student discontent always negatively impacts students, staff, administrators, and institutions' aims. Nigerian schools, according to Lawal (2003): Loss of life; destruction of public and private property; disruption of academic programs; loss of government income; diverting government focus from other critical areas of the economy Obianyo (2003), heads of institutions have been accused of the alarming growth in student unrest, rioting, and vandalism, particularly at the secondary level since they lack the skills and understanding necessary to control student discontent.

## 2.3 Nigerian Socio-Political Contexts

Developing a country is crucial to its long-term viability and progress. When a nation can offer a quality of life for its citizens, it is considered developed. Although Nigeria possesses a wealth of people, material, and natural resources, the country has struggled for decades to grow.

Health care, education, housing, and other essential services are part of what Naomi (1995) considers to be standard definitions of development when enhancing the community's individual and communal well-being (Naomi, 1995).

As defined, Chrisman (1984) defines development as societal growth that aims to enhance people's well-being by collaborative efforts including a wide range of sectors, corporations, and other social groupings. While economic growth is an essential component of development, it is necessary to consider that social, political, and technological factors all have a role.

Due to the country's sociopolitical backdrop being negatively influenced by a lack of solid administration. Development is a fantasy if it is not accompanied by sound government. This is due to the country's poor leadership. Most of our leaders lack a strong sense of commitment to growth.

Another obstacle to growth is a high degree of corruption and indiscipline. The government in Nigeria is run by corrupt officials who have turned it into a tool for profit accumulation rather than a means to advance the people's interests in the hands of a government that has been infiltrated by organized crime; even the best-laid plans will fall short of their full potential (Mimiko, 1998). When one is present, the other suffers. Corruption and progress are incompatible, and the two cannot coexist.

For a long time, poverty and unemployment have been vital development issues in Nigeria. According to Obadan and Odusola (2001), unemployment in Nigeria peaked in the 1980s and has been steadily rising ever since. In 2008, 15% of the nation's workforce was jobless, and by 2011, that proportion had increased to 20% (Lamido, 2013). According to Akanda and Okuwa (2009), 40-60% of jobless Nigerians are aged 15-25, whereas Rotimi (2011) says the age range is 18-45. unemployed are destitute, and most life into the old life. Most of those paraded before the media as members of criminal gangs are in this age range. Terrorist organizations employ this age range for most of their suicide bombers. Most Niger Delta militants are also in this age range. In reality, most of these youths would not engage in criminal activity if they had other sources of income or were working.

Worst of all, some of those implicated in these acts are educated. The worst thing that can happen to a country is for educated people to be involved in crimes, which are cybercrimes and advance-fee scams. Educated people are engaged in non-violent crimes that are worse in scale and scope than violent crimes. They have higher implications and effects on the country's socio-economic processes and administration. The illiterate young are enraged and frustrated by the educated and resort to more aggressive measures to survive. Poverty and unemployment have destroyed practically every country on the globe. However, as Akande and Okuwa (2009) point out, the rising number of jobless youngsters strolling the streets of Nigeria captures the unemployment crisis. It is perceived in the informal economy as underemployment, dropping real wages, diminished incentives, lower private investment in all sectors of the economy, and a decline in the quality of education and training provided to residents throughout the country. This leads to poverty and increased criminality. In other words, unemployment in Nigeria poses social, economic, political, psychological, and security difficulties.

### 2.3.1    Special Anti-Robbery Squad (SARS)

In Lagos State, the special anti-robbery squad (SARS) was formed in 1992 in response to the actions of the infamous armed robbers in the region (Malumfashi, 2020). The Nigerian Police Force had a section dedicated to combating violent crimes, and it was called the Special Response Unit (SRU). To carry out covert operations against armed robbers, it was created as a faceless police squad (Oloyede & Elega, 2019). A quarter of a century after it was founded, the unit had expanded throughout the nation. Some of the unit's unruly qualities were caused by dispersal and lack of organization. The term police conjure up images of bribery and corruption in the minds of most Nigerians. Instead of fighting crime, SARS has long been suspected of being plagued with it as a result of this (Adepetun, 2020). When the police used all of their authority to exploit individuals and brutalize those who didn't cooperate with their financial extortion, it was discovered that the purpose for establishing SARS had been overcome. Numerous Nigerians complained about police misconduct, but nothing was done about it despite the outcries of hundreds of people who had been harmed. Armed rapes, extrajudicial murders, and other acts of torture, like those in Delta State, were common place among SARS agents. They were not held accountable (George, 2020).

## 2.4  Social Media's Impact on #EndSARS Protest

Social media has ushered in a new era of communication since its beginnings in the 1990s (Cammerts, 2015). Today, billions of people across the globe are digitally connected and active on the internet. Even emerging markets have used various methods to keep their populations abreast of technology advancements in their fields (Obaid, 2020). The global economy has become even more technologically reliant because of the current worldwide epidemic, covid-19, which has caused people to remain inside and rely only on technology for communication. People who were previously isolated and underprivileged now have access to communication networks thanks to social media in today's volatile political climate. It has also promoted the sharing and exchange of ideas, views over global distances via social media (Ohia & Salawu, 2020).

Additionally, social media has given users the chance to get information from others rather than depending on what they already know through the media. Nigerians have seen a massive rise in social media use due to the increased use of web 2.0 technologies over the last two years. This increased use has helped many social movements, like the #EndSARS protest, which received worldwide attention (Aderinto, Ogunleye, Ojedokun & 2021). Over the last decade in Nigeria, social movements have played a critical role. A group of activists or anti-government actors frequently initiates social activities to voice complaints, interests, criticisms and suggested solutions to particular recognized issues via collective actions (Obaid, 2020). These crucial players have traditionally been involved in social movements and led and organized demonstrations. Although social media have effectively supplanted the pre-existing mobilization mechanisms, it has become an essential instrument for managing and coordinating worldwide social movements (Aderinto, Ogunleye, Ojedokun & 2021).

It's not only the mobilization of money affected by social media; it's also the reorganization of relationships inside social movements. It also made it possible to restructure public spheres and completely alter how knowledge and communication move in society. Social network structures have shifted dramatically as a result of this development. (Obaid, 2020) said that social media has two significant effects on social movements. The first is that social media speeds up recruiting, mobilization, communication, and information distribution, opening up new mobilization venues that weren't previously available via conventional mobilization tactics. Several social media apps allow billions of people to connect worldwide (Ohia & Salawu, 2020).

People of various races, ethnicities, and identities have used social media as a platform to communicate, develop, and gather emotions or sentiments, resulting in a new kind of virtual collective identity on which social movements have bowed out of their demonstrations (Cammerts, 2015). Most importantly, social networking has made it possible to validate the information instantly, overriding the system's inherent bias in the news media. In the wake of this change, social movements can now spread a positive narrative and evoke empathy among their supporters. Take the social media campaign to EndSARS as an example (Mariam & Olabamiji, 2020). Disappointment and dissatisfaction over the weak leadership in government and structural inequalities in various parts of Nigeria were expressed through social media. People were fed up with the educational system, poor health care, insufficient health care, terrible roads, and a lack of basic infrastructure, to name a few problems in their communities (Adewole & Osabuohien 2007).

## 2.5 EndSARS Protest

The EndSARS movement started when convener Segun Awosanya published a petition (Elega & Oloyede, 2019). EndSARS began as a call for the dissolution of the Special Anti-Robbery Squad (SARS), a unit of the Nigerian police known for its brutality and violations of human rights. The EndSARS protest responded to the widespread dissatisfaction of Nigerians with the police agency's service delivery system. The campaign moved many Nigerians because SARS embodied everything they disliked about Nigeria's government (Africa Center for Strategic Studies, 2020). According to an investigation, Nigerians have poor trust in the police since many have been compelled to bribe officers in the past. Public outrage was sparked by a social media post regarding the death of a young man in Delta State by SARS agents in October 2020. Several young Nigerians immediately took to social media to voice their displeasure with SARS. Their own experiences with IRS quickly developed into a movement that offered Nigerians fed up with the extortion, injustice, and violence of the infamous police force an outlet to express their displeasure. The Federal Government of Nigeria soon announced the abolition of SARS due to the media attention the demonstrations had generated. It didn't help that the IGP had made a similar statement in December 2017 about SARS "stop and search" operations after many complaints of harassment reached the Nigerian Police since the cancellation did not satisfy the protesters. In 2018, the IGP publicly rescinded the prohibition, demonstrating the inadequacy of previous rulings (Fasunwon, 2019). As a result, the Nigerian people lost faith in the police. The EndSARS demonstrations drew support from officials across the globe, including United Nations Secretary-General António Guterres and United States Secretary of State Mike Pompeo (Uwazuruike, 2020). The hashtag was also used by well-known figures and celebrities, including former US Secretary of State Hillary Clinton and current US Vice President Joe Biden, as well as Arsenal player Mesut Ozil, heavyweight boxing champion Anthony Joshua, and American musician Kanye West (Larnyoh, 2020; Uwazuruike, 2020). In October, the hashtag EndSARS was the most trending subject, thanks to Twitter CEO Jack Dorsey's introduction of a special emoji for the cause (Uwazuruike, 2020).

This research of global demonstrations from 2006 to 2013 showed that a large proportion of protests was against neoliberal reforms such as privatization of public businesses, full-scale deregulation of public utilities, and the adoption in Third World nations, especially of different opponents The International Monetary Fund (IMF) and the World Bank, whose structural adjustment prescriptions have taken away a country's sovereignty, were criticized by protestors in Third World nations who wanted a strong state capable of resisting their demands, according to Rao (2010). Activists protesting against neoliberal reforms called for more government funding for social services and public utilities and reduced corruption. This was especially true in the petroleum industry, where shylock and oligarchic marketers ruled.

Hence, the purpose of protest is to remind people in authority of the daily injustices that disadvantaged groups face. SARS is known for its brutality, but human rights violations by security personnel are not limited to the NPF, according to many theories as to why the EndSARS protest has erupted. Amnesty International released a report in 2015 detailing Nigerian military abuses. Security forces had violated human rights since Nigeria's military dictatorships when security services like the police were created to preserve military governments rather than protect and serve people. Although the democratic government was established in 1999, this mentality persists. Intimidate peaceful demonstrators and arrest journalists and other government opponents using police and military troops (Nkasi, 2020).

In addition, the lack of accountability for SARS and other security forces' human rights violations is a significant driving force for the EndSARS demonstrations. The police hierarchy often protects security agents who violate human rights, and as a result, victims are reluctant to come forward for fear of retaliation. Since the National Assembly approved the Anti-Torture Act in 2017, according to an Amnesty International study released in June 2020, no SARS operatives have been prosecuted (Amnesty International, 2020). The Police Act, which established the NPF, is also out-of-date. The law, which wan 1943, gives cops a great deal of latitude in their duties. These empower police to conduct warrantless searches and make arrests based on mere suspicion of guilt or even of criminal intent. SARS has been accused of seizing victims' phones without a warrant and arresting people for their tattoos or dreadlocks, which is at the heart of many accusations.

Many academics believe that social media has played an important role in recent demonstrations (Silva, 2015). Due to the ease with which it provides access to a vast number of connections, creates collective identities, and serves as a platform for information dissemination, social media may be an essential tool in mobilizing (Arriagada, Scherman and Valenzuela, 2012). Social networking does not always generate new forms of protest or change conventional organizations fundamentally, as several scholars have quickly pointed out. A recent study suggests that activism should not be confined to the internet. This means that traditional modes of protest are supported or made more accessible by social media. Additionally, protesters utilized social media to coordinate their actions and denounce police brutality; they also published a counter-narrative to the official government narrative via platforms like Twitter and Facebook.

## 2.6    Theoretical Context

Although the research study adopted the inductive research approach, the socio-materiality theory is in line with its findings and therefore was used to inform the theoretical elaboration done in the study. Theories are systems of assumptions, principles, and linkages proposed to explain a particular set of occurrences (Bates, 2005). On the other hand, Gray (2013) found that theory links concepts about the phenomenon, offering a cohesive framework to explain or foresee happenings. Consequently, the socio-materiality theory was used as the framework to interpret the data collected in the cause of the study. The socio-materiality theory describes how social practice and materiality interact in organizations (Orlikowski, 2007; Orlikowski, 2010). The theory was propounded based on the separation or duality assumptions used by scholars to describe and explain how information technology impacted the achievement of organizational goals (Orlikowski, 2007; Orlikowski & Scott 2008). The socio-materiality theory proposes that information systems, everyday life realities, and the contexts where the information systems and everyday life realities intersect are embedded. The theory draws scholars' attention to the embeddedness of human actions and experiences, information systems, and contexts (Orlikowski, 2007; Orlikowski & Scott 2008).

# Chapter 3

# Research Methodology

## 3.1 Introduction

The chapter deals with the study's philosophical perspective, the method and strategy used to collect and analyze data, and the presentation of the findings. The research approach and procedures utilized in the research process were explained in the chapter. There is also a detailed explanation to justify the options used to carry out the research study. The scope and limitations of the chosen study design, as well as its position within existing information system methodological assumptions, are briefly defined in the chapter. In this chapter, I also explained the study's philosophical foundation. In the information systems discipline, the philosophical foundation explains the philosophical notions underpinning researchers' assumptions about the realities, values, concepts, and the nature of knowledge in scientific inquiries. Research philosophies allow the explication of how contextual factors within the research contexts are handled, the role of the researcher, and the nature of knowledge. It also will enable researchers to justify their choice of research method, and the implication of the research method chosen on the strategy of inquiry and the samples were selected to be studied. Consequently, the segment also includes my explanations of the research method (the case study research method) that I adopted in the study. I also explained the sample and the sampling technique I adopted in the segment. As a result, the part covers the study's research design, approach, philosophy, methodology, sample and sampling technique, data collection and analysis techniques, and ethical considerations for the research study.

## 3.2 Research Design and Approach

According to Labaree (2006), research design denotes the entire strategy adopted by a researcher to integrate the different components of the study logically and coherent. It enables the researcher to answer the research question(s) effectively and serves as guide to data collection, measurement, and analysis (Creswell, 2014). There are three major types of research design: quantitative design, qualitative research design and mixed method design (Creswell, 2014). The study adopted qualitative research design and consequently, was informed by the qualitative data collected on how study participants perceived their everyday lived life as individuals and groups, and how they make meaning from their everyday life experiences. So, in this study qualitative research design was adopted to examine how social media impacted the EndSARS protests in Lagos, Nigeria. I observed in real-time the EndSARS protests in Lagos, Nigeria. I also participated in the dialogues and communication on the issues that built up to the EndSARS protests. In the course of the study therefore, the initial experiences I had before the study came to bear when I started to conduct the interviews for the study. I interviewed participants that were involved in the EndSARS protests in real-time and on social media.

 The research approach adopted for the study is the inductive research approach. The tenets underlying the inductive approach align with those of the interpretivist philosophy. Inductive research approach allowed the data collected in the course of the study to be collected without using existing formal theories. The inductive approach allows the researcher to utilize real-life events as examples to present subjective reasoning. Because the inductive research approach was used to inform the study, I began by collecting qualitative data from which I discovered a pattern for coding study participants' responses into meaningful variables with relationships (Chen, et al. 2018). The inductive approach adopted in the study is similar to that used by Utulu (2019) and Utulu & Ngwenyama (2019) approach which is rooted in Charmaz's (2014) variation of the inductive research approach.

## 3.3 Philosophical Foundation of the Study

Philosophical assumptions in the social sciences provide grounds for developing assumptions on how the researcher views the realities surrounding the phenomena s/he is studying (Creswell & Poth, 2017). For example, the philosophical foundation of a research study enables the researcher to express their understanding of reality (ontology), how they understand and study reality (epistemology), the inquirer's value-stance (axiology), and the implications of all these stances on the processes followed to conduct the inquiry (methodology) (Cresswell & Poth, 2017). There are three primary research philosophical paradigms used for social sciences-based inquiries, including in the information systems discipline. The philosophical paradigms are positivism, interpretivism, and Critical Realism (Burrell & Morgan, 2006; Orlikowski & Baroudi, 1991). The interpretivism was adopted in this study. The interpretivist philosophy is a research paradigm that is based on the assumption that social reality is contained in individual subjective experiences of the external world (Walsham, 2006). The interpretive philosophy is often used synonymously and loosely with the qualitative research method. However, the two concepts are different given that interpretivism is shaped by social contexts and human subjective experiences ontology (Utulu & Ngwenyama, 2017).

 According to Walsham (1993), there is no correct or incorrect theory in interpretive research. Theories are chosen and used by interpretivist researchers based on researchers' motivation. Researchers derive their conceptual framework by critically examining the phenomenon been investigated. Interpretive research is often accomplished by observation and interpretation of interview data collected. Walsham (2005) postulates that the interpretive mode of study does not build a research theory but instead evaluates and refines existing theoretical notions. Reality is relative and multiple (Hudson and Ozanne, 1988). According to  Uduma & Waribugo (2015), many interpretive truth depends on other systems for meanings; this makes it difficult to interpret in terms of reality. Knowledge acquired from interpretive research is socially constructed rather than been seen as being objective and perceived (Boer, 2005). Given that the study adopted the interpretivist philosophy, it aims to provide understanding on how social actors socially constructed the realities that made social media impact on the EndSARS protest in Lagos, Nigeria. The study was informed by the notion that protesters, the contexts where they protested in Lagos, and the social media platforms used as communication tools were all socially constructed. In other words, they were historical and subjective entities. Consequently, discussions done during the interviews were purposefully broad and generic to allow participants to build meaning and provide their views of the EndSARS protests in Lagos, Nigeria. The interpretive philosophy enabled me to see the phenomena surrounding the EndSARS protests from the participants' perspectives as noted by Creswell's (2014). I also adopted Klein and Myers' (1999) principles of suspicion and dual interpretations to make more constructive meanings from the data collected during the course of the study. The study's primary goal of identifying how social forces promoted the EndSARS protests in Lagos, Nigeria was reached.

## 3.4 Research Method

A research method is a strategy for putting the research plan into action and are well-thought-out, scientific, and value-free (Saunders, 2017). The case study research method was used for the study. A case study is a research strategy used to get a thorough, comprehensive understanding of a challenging issue in its real-world context. It is a well-established research approach frequently used in various subjects, particularly the social sciences. A case study enables an in-depth analysis of a small sample size and even allows the study of a single sample in a single case study scenario (Saunders et al., 2007). The case study is the most versatile research design because it enables the researcher to retain the holistic aspects of real-life happenings while investigating empirical events. Case study research is divided into single case studies and multiple case studies (Gable, 1994). It could be quantitative case study Yin (2009) or qualitative case study Lee & Liebenau (1997). Benbasat et al. (1987) exposed the problem the information systems community suffers due to the use of elaborate research methods. They posited that the information systems community is "characterized by constant technological change and innovation. IS researchers, therefore, often find themselves trailing behind practitioners in proposing changes or in evaluating methods for developing new systems [and] usually learn by studying the innovations put in place by practitioners, rather than by providing the initial wisdom for these novel ideas (p. 370)." It is a more suitable option for carrying out study credence to participants' perspectives and consequently, aligns with the philosophy underlying the interpretive philosophy as propounded by Myers & Klein (1999) and the inductive research approach.

I adopted the case study research method for the study. Youths in several states in Nigeria, particularly in western Nigeria, including Osun, Oyo, Ogun, Ondo, and the Ekiti States, participated in the EndSARS protests within their localities. I, however, chose Lagos State as the case study state to carry out the study. This is a form of single case study and is consistent with the following position about a single case study. The EndSARS protests in Lagos State are unique and incomparable to those in other western Nigeria States in terms of traditional electronic press and social media coverage. It is also easy to access Lagos State from Yola, Adamawa State, given that there are direct local flights from Yola to Lagos. As at the time of the study, other states in western Nigeria are accessible directly from Yola. This made it easy to have access to those that participated in the study and carry out interviews and observations done during the cause of the study.

### 3.5 Research Population, Sampling Technique, and Sample Size

In research, including information systems research, a population can be defined as a large group of people, institutions, items, and other entities that share standard features and meets the parameters set by the researcher for the inquiry is plans to carry out (Ritchie et al., 2013). The shared qualities set a population apart from other competing potential population that could be studied. In the social sciences and the information systems discipline in particular, looking at the entire population is nearly impossible for a researcher. According to Ritchie et al. (2013), in research, there are two composed groups of the population, namely, target population, which is the entire group of people or objects to which the researcher wishes to generalize the study findings and accessible population which is the portion of the population to which the researcher has reasonable access; maybe a subset of the target population. Consequently, the study population comprises all the youths involved in the EndSARS protests in Southwest, Nigeria. To come up with the sample population of the study, the multilevel sampling was adopted. First, the state to be studied among the six Southwestern states was sampled, followed by the youths who participated in the EndSARS protests. The Lagos state was chosen as the sample state using the convenient sample technique while the purposive sampling technique and the snowballing sample technique were used to sample the youths that participated in the study. The convenient sampling technique, purposive sampling technique and snowballing technique are non-probability sampling approach in which the sample is chosen from the population based on the ease with which the sample can be reached and their direct relevance to the subject(s) of the study (Saunders et al., 2007).

### 3.5.1 Size of Sampled Participants

The study sample size comprises two groups, namely protesters and activists. Activists are organizers of protests, and they begin agitations and recruit protesters either face-to-face or on social media. The protesters are part of the civilian population whose everyday lives are impacted by the subjects, usually on injustice, bad governance, police brutality, and unemployment. The study sample comprises twenty-five (25) protesters and five (5) activists. There is no unique suggestion on the sample size needed for single case studies (Guest et al. 2006). The sample size depends on the size of the project, the type of data collection, and how the themes are analyzed and reported (Fugard & Potts, 2015). The important thing about the number of participants allowed in single case studies is the in-depth analysis of the research problem and the collection of relevant in-depth data on the research problem (Saunders et al. 2007). The table below shows the study's sample size and the groups that comprise the study.

| Category of participant | Number of interviews |
|---|---|
| Protesters | 25 |
| Activist | 5 |
| Total | 30 |

**Table 3.1 Number of Interviews held during the Study**

## 3.6    Method of Data Collection

The unstructured interview technique and participant observation were adopted as the techniques for data collection in the study. I spent between six- and eight-months observing issues relating to the EndSARS protest on my social media handles including, Twitter, Instagram, Snapchat, Facebook, WhatsApp, and YouTube. The formats of the information available on the social media through which I observed events revolving around the EndSARS protests are texts, video, pictures, audio, and audio-visual formats. Together with the interviews held with human participants, these sources and forms of information constituted the primary research data sources. However, there are insights in the literature on interviews in social science, including information systems research. Schultze & Avital (2011) argue that discussions occur when a participant asks another participant constructive question about set themes to generate data on the themes. Interviews can be structured, semi-structured, or unstructured (Mojtahed et al., 2014; Mingers, 2003). The type of interview adopted in this study is the unstructured interview. Unstructured interviews are interviews carried out by researchers in which the questions are not prearranged or derived from existing theories or theoretical perspectives (Nandhakumar & Jones, 1997). The in-depth unstructured interview is the opposite of the structured interview, which is arranged and dependent on questions derived from existing theories or theoretical perspectives. The structured interview looks to raise questions and elicit a response that aligns with the theory's insights that drive the research for which the interview is done. Unstructured interviews align with the requirements of research studies based on the inductive research approach. This is given that the studies are not driven by any existing formal theory or theoretical insights or based on a posteriori knowledge. I decided to adopt the in-depth unstructured interview because of the need to develop more specific and contextual findings in the study given the novel nature of the event, the EndSARS protests under investigation. Thirty participants were interviewed during interview sections that lasted between thirty-five and sixty minutes in the cause of the study. Some of the unstructured interviews were recorded using my mobile phone, while some were documented in my field note. Participants that permitted recording had the unstructured interviews held with them recorded. Those who did not allow recording devices had their interviews recorded in my field note.

### 3.7. Data Analysis Technique

The data analysis technique adopted in the study is the thematic data analysis technique; according to Braun & Clark (2006), the thematic data analysis technique groups qualitative data derived during interviews or observations recorded in field notes into themes. Using illustrative quotes from participants, this research approach helps researchers better comprehend participants' experiences when phenomena emerge (Silverman, 2011). This sort of analysis increases openness by revealing the nature of data and how it is represented in the interconnected context of the stakeholders, as interviews would reveal repeating patterns of the meaning of the research topic (Neuendorf, 2018). Flexibility, accessibility to the conduct, and findings for qualitative researchers and the general public are all benefits of theme analysis (Braun and Clark, 2006).

I adopted Braun and Clarke (2006) are thematic data analysis technique steps. The steps include:

1. Familiarizing myself with the data: This is the initial stage of the investigation. During this phase, I read and reread the data to get acquainted with it.
2. I generated initial codes: I identified patterns and relevant codes in line with the purpose and objectives of the study. Consequently, codes like unemployment, SARS brutality, bad governance, etc., emerge at this stage.

3. I then searched for the themes in the data in the transcribed interview scripts and field notes.

4. I reviewed the themes and evaluated for coherent and consistency. Patton (2002, p 465) states that the themes should have "internal homogeneity and outward heterogeneity." This was taken into account while assessing the themes.

5. I defined identified themes. Once the themes were determined, they were tagged and defined. They were characterized by what they portrayed in the context of the study.

I wrote the research findings and theoretical elaboration and then refined the introductory chapter, literature review, and research methodology chapter.

### 3.7 Research Process

1. Step 1: I began the study with participant observation of events during the EndSARS demonstration in Lagos, Nigeria, since I had been engaged in the research surroundings for months. Most protesters knew who I was and what I was doing by this time. This familiarity helped me engage with them. During my participatory observation, most participants utilized a digital device to communicate. So I created the following research question: How did social media impact street protest during the EndSARS protest in Lagos, Nigeria?

2. Step 2: I used snowball sampling. My first interview was with a protester. Then I conducted interviews with other protesters and activists. I interviewed 30 people in all, 5 EndSARS activists and 25 protesters. Some interviewees declined to have their voices recorded. Instead, they preferred we have a dialogue and noted down points stated. I used an iPhone 6 Plus with zoom for the few that permitted it. I kept a notebook of my observations and examined them before and after interviews. Interviews lasted a month and averaged 30-40 minutes. I spent two weeks verifying replies with study participants. Step 2 took me two months in total.

3. Step 3: I examined the in-depth interviews and participant observation data. I utilized ATLAS. ti to analyze data and capture results in my field diary. I went back to sure study participants clarified some of their confusing claims to me. For specific replies, I verified with research participants. During this stage, I developed the research model for study two.

4. Step 4: I typed up research two and thought about the knowledge gaps identified. My thoughts helped me understand why the observed gaps are consistently ignored in the literature.

### 3.8    Ethical Consideration

EndSARS protestors and activists made up the majority of the sample group. Considerations for voluntary participation, the use of personal data, and public disclosure by research participants of sensitive information have been taken into account on an ethical level. Only a small percentage of participants in the research had the option to seek access to "sensitive" data. Despite this, participants were fully informed of the risks of using sensitive information they had voluntarily submitted. As a result, they understood how much information regarding the protest and the problems experienced in Lagos during the EndSARS protest in Nigeria would be needed for the research. Confidentiality, data ownership, and consideration of data that may be made public as a study output were all established and agreed upon by the interviewees and myself. The agreements included study participants as well. According to a professor from the American University of Nigeria (AUN), my methodology and ethical claims were thoroughly examined to ensure that they were free of ethical concerns, given the kind of research study I intended to conduct.

# Chapter 4

# Presentation and Discussion of Findings

## 4.0 Research Findings

This chapter provides the study's findings, which were derived from the data gathered through in-depth unstructured interviews with activists and protesters and my observation as a participant on social media. The results explain the factors that come to bear in the ways social media impacted the EndSARS protest in Lagos, Nigeria. Given that social media played a prominent role during the EndSARS protests, it became practical to ask and provide answers to how social media impacted the EndSARS protests in Lagos, Nigeria. The question becomes necessary because of the apparent mix between real-time activities in the streets and the crowds of information transferred on social media by the youths involved. The data presented in this research report segment were used to provide insights on the embeddedness of the Nigerian socio-political context, the everyday life experiences of those involved in the EndSARS protests, and the social media. Thirty in-depth unstructured interviews were held with activists and protesters to use insights provided through them to explain how social media impacted the EndSARS protests. I also linked the data to the research questions, theoretical insights, and two propositions in this segment. The part is written in two main sections: presenting research findings and theoretical elaboration.

The demography of the study participants is as follows. Five main activists that have been involved in using social media to agitate against unfavorable government policies and bad governance for about two to five years participated in the study. Consequently, they have what one may describe as good enough experience in agitation against the government. The five activists have university qualifications and were within the age range of 35-45 years. The twenty-five protesters that participated in the study included seven female protesters and eighteen male protesters. Their age range was between the ages range of 25-45.

Most of them were learned and had at least secondary school educational qualifications.

**Table 4.1: Table showing the study participants details:**

| Participants | Role played | Gender | Educational qualification | Age-range |
|---|---|---|---|---|
| 1 | Activist | Male | MSc Sociology | 30 |
| 2 | Activist | Male | MSc Political Science | 35 |
| 3 | Activist | Female | MSc Management | 40 |
| 4 | Activist | Male | MSc international pol. | 38 |
| 5 | Activist | Female | BSc Economics | 25 |
| 6 | Protester | Female | BSc Accounting | 31 |
| 7 | Protester | Male | HND business | 26 |
| 8 | Protester | Male | OND computer sci. | 30 |
| 9 | Protester | Male | Sec. School Cert. | 22 |
| 10 | Protester | Male | BSc Engineering | 25 |
| 11 | Protester | Male | BSc Mathematics | 29 |
| 12 | Protester | Male | BSc Economics | 30 |
| 13 | Protester | Male | BSc Banking & Fin. | 32 |
| 14 | Protester | Male | Sec. School Cert. | 20 |
| 15 | Protester | Male | Sec School Cert | 22 |
| 16 | Protester | Male | OND Accounting | 24 |
| 17 | Protester | Male | HND Business Edu. | 30 |
| 18 | Protester | Male | BSc computer sci. | 29 |
| 19 | Protester | Male | BSc info tech. | 31 |
| 20 | Protester | Female | BSc Geography | 32 |
| 21 | Protester | Female | Sec. School cert. | 20 |
| 22 | Protester | Female | OND info system | 23 |
| 23 | Protester | Male | HND Business | 30 |
| 24 | Protester | Male | Sec. School cert. | 29 |
| 25 | Protester | Male | HND computer | 28 |
| 26 | Protester | Female | Sec. School cert. | 25 |
| 27 | Protester | Male | BSc. Agriculture | 28 |
| 28 | Protester | Female | BSc Home Economics | 27 |
| 29 | Protester | Male | Sec. School cert. | 24 |
| 30 | Protester | Female | BSc English | 28 |

**Presentation of Study Findings**

Interview transcripts analyzed using the thematic analysis technique revealed a total of 17 codes. Codes sharing similar underlining concepts were grouped into 10 categories from which 6 themes emerged, as presented in the table below:

**Table 4.2: Table showing the themes and categories:**

| Themes | Sub-categories | Code |
|---|---|---|
| Bad governance | Grievances<br><br>Basic amenities not provided | Lack of basic amenities |
| | | Lack of voice and accountability |
| | | Grievances and brutality of protesters most especially by government agents |
| SARS brutality | Enlightened through social media<br><br>No proper identification | Victimization of victims falsely |
| | | Using of force to collect money on the checkpoints. |
| | | Fear of the unknown due to the search and stop mostly at night on the highway. |
| | | Commit a crime and still won't be charged cause of identification issues most SARS personnel wear personal clothes, not uniforms. |
| Misuse of Public Power | Falsely accused<br><br>Above the law | Failure to bring perpetrators to book |
| | | The use of torture leads to the victim's death, missing without proper investigation. |
| | | No proper accountability of victims when arrested |
| | | False accusation of victims when the crime is not committed by the accused. |
| Corruption | Victimization | Increased payment to uniform personnel to eradicate corruption |
| | | Make sure that search and stop officers have microphone connected uniform that has cameras |

| Unemployment | Caste system (a class structure that is determined by birth | Poor educational system leading to unemployment |
| | | Lack of proper jobs that can solve basic needs resulting in hunger |
| | Hunger | |
| Insecurity | Fear of the unknown | Safety of citizens when carrying out daily routine. |
| | | High rate of kidnapping in the country. |

Six main themes emerged from the analysis as presented in the table above.

## 4.1 Social Media, Nigerian Socio-political Context, and the EndSARS Protests

The study began to find out how social media impacted the EndSARS protests. The term EndSARS was coined from the protests held to disband (End) a special force in the Nigerian Police Force, namely the Special Anti-Robbery Squad (SARS). This was important because most scholars tend to explain how social media impact protests in contemporary societies from the dualist and information technology determinism perspectives. The study aims to provide an alternative explanation. Consequently, the first factor identified during the investigation that led to the EndSARS protests is the problematic Nigerian socio-political context. In the study context, the Nigerian socio-political contexts denote occurrences in societies across Nigeria, including bad governance, misuse of public power, corruption, SARS brutality, unemployment, and insecurity. Socio denotes social contexts defined by norms, values, cultures, and the interactions among and within people. Political parties deal with polity, political space, rules and regulations, laws, and general expectations to keep society habitable, peaceful, and just. Therefore, socio-political deals with events (bad governance, misuse of public power, corruption, SARS brutality, unemployment, and insecurity) made those who participated in the EndSARS protests use social media to interact and share information. The social media-based interaction and information sharing led to the buildup of organized civil protests against SARS. My observation based on information shared on social media indicates that the relationship between the SARS and citizens in the southwestern part of Nigeria, particularly Lagos, was problematic. The SARS is a Nigerian police force formed in late 1992 to deal with robberies, car thefts, kidnappings, cattle rustling, and gun offenses. However, the SARS operative's mandate was to protect the life and property of citizens became a problem to those they were supposed to protect. This raises questions regarding how social media impacted the EndSARS protests. The findings of the study regarding this question are presented below.

### 4.1.1 Bad governance

Bad governance in context results from decision-making, connecting those who rule and those who are governed. Bad government includes dishonesty, the passage of unjust policies, lack of transparency and accountability, arbitrary policymaking, and defrauding the governed. In 2015, the ruling party promised to bring about "change" for the country's development which boosted the expectations of many Nigerians, most especially the youths. Seeing themselves dominated by the old, the country's young felt excluded from policymaking. They wanted their opinions to be heard and a functioning system, but that was a long way off.

According to interviews conducted, the participants had different views on bad governance, but most believed that bad governance is the root of every societal problem. Participant 8 thought that the government, through its leadership skills, gave room to breeding the act exhibited by the SARS units. *"Am enraged by the absence of basic facilities, enraged by the lack of power, poor roads, a broken academic schedule, and slow economic growth of the country. I wake up thinking of all these basic stuff that makes life easy, which I can't afford; that's the case of so many SARS operatives; they want to have all of those basic amenities, so they* *do what they must do; to survive." (Participant 8)*. However, Participant 5 said that *"#EndSARS is more than just a protest. It's about #Nigerians voicing out (soro soke) on bad governance, especially the poor working class that can't be heard." (*Participant *5)* during the interview section, I observed that both participants 8 and 5 were blaming the government due to their interest or experience with the things not provided to them by the government. Participant 3 believed the reason people took to the street of Lagos to protest is that the same government has time without number announced a structural change to SARS. Still, the alleged human rights violations and exploitation continued. For instance," *in December 2017, the Inspector General of Police (IGP) announced that SARS had been banned from conducting stop and search operations following several reports of harassment. This ban was publicly re-announced by the IGP in 2018 and 2020, reflecting the ineffectiveness of previous orders. In light of past practices and disappointments, protestors added to their list of demands, calling for the compensation of victims of SARS brutality, retraining of police officers, and trials of indicted SARS officials.*" (*Participant 3)*

The participant that drew my attention the most was participant 4; he joined the protest because he was grieved. *"I lost my brother due to poor infrastructure, hospitals having no equipped modern facilities to care for its citizens. I had to protest because I wanted to change. If everything were working perfectly so, many would have been different." (Participant 4)*

Based on observation, all the participants quoted above showed signs of deep concern when expressing their opinions on bad governance. Their reaction and rage while discussing the bad governance showed their grievance on why they had to participate in the EndSARS street protest in Lagos, Nigeria. The references were made to the promise of a better life at the dawn of democracy in 2015 failed to materialize. Social media and protest were the only way they felt would draw attention to the government to improve their lives. Once a system is wrong, it leads to so many harmful policies the government needs to restructure the different units of the SARS team to provide basic amenities to avoid the reoccurrence of yet another protest.

## 4.1.2 SARS brutality

 SARS brutality is seen as an act of using excessive force by a SARS police officer against an individual; most times, such volition results in the death of the charged victim. For instance, participant 2, an online marketer, said, *"When I was around thirteen, a SARS officer threatened to slap me. As an adult woman, I've learned to fear SARS-like everyone else. I lock my vehicle doors as soon as I get in, not because of thieves, but because too many SARS have barged in to extort me. On odd days, a SARS unlocks the vehicle from inside before I can wind them up. They only leave once I pay them. But even my worst days were better than hundreds of others." (Participant 2).* I understood that her case was a youth resulting from what had transpired between her and a SARS officer while growing up. I even went as far as to ask why the policeman tried to slap her, but she waved the question; the stigma grew with her. It became worst as she encountered more horrifying stories on social media about people victimized by SARS operatives; she still went further to say that *"I first heard about the SARS brutality on Twitter through the Hashtag #EndSARS; initially, I didn't take it seriously until I came across a negative content (video) of how the SARS team brutalized on Twitter which I followed the comments and viewed the comprehensive documentary that made me cry." (participant 2).* I noticed that the fear of the SARS operatives already embedded in her mind to her every uniform personnel is terrible. However, I tried to give her reasons to see that not all uniform personnel are awful; all effort proved abortive her mind was made up, and nothing can change it. Participant 6, who is a business owner that sells clothes in Lawansi, had a similar story of how the SARS unit brutalized; his ordeal was somehow different from participant 2; he narrated his story with rage and anger according to him,

*"Keeping late nights because of the nature of my work, it happened on a faithful day that I was supposed to drop my friend that is living in Aguda 40 brown road,*

*and I was coming from Lawansi approaching Aguda junction there was a stop and search going on was asked to stop, open my boot and all was found clean but still was asked to drop something for the boys. I tried to tell the SARS officer I had nothing on me, but to my greatest surprise, he demanded a transfer or delayed me from going to where I was going. My friend and I tried to raise voices at the officer, who, in turn, because of the level of drinks (alcohol), had threatened to shoot us if we challenged him. Any time I think of that situation, I feel sorry for the countless people who have been threatened like that; maybe they might not be as lucky as I was." (participant 6)*

I tried to calm participant 6 down, seeing his anger in explaining his encounter and showing how SARS operative pushed people. However, participant 1 had a different encounter due to a false identity. He narrated his ordeal, which moved me to feel sorry for him.

*"I was coming back from my usual Friday club groove me, and my friend at about 1 am because that like a routine for us going out to recreate after a busy week, it happens on this faithful, there was a robbery incident going about the area we happened to be driving back, we were blocked and arrested by the SARS operatives. We went further to show our identification cards but to our greatest surprise, were handcuffed and taken to the police station; on getting there, we were stripped and charged as robbers put in a cell and beaten up as criminals our families the next day had to come to bail us out after a huge sum was collected from us. SARS, like other Nigerian police organizations, regularly detains people for years without trial that I got to know about when we eventually came out."* (Participant 1)

The unit should be overhauled entirely based on observation, and the team should be rebuilt. That is when the government steps in, recruiting qualified personnel and providing them with a sufficient take-home income and staff an administering office that will be inspecting the SARS units. The government must enact legislation to protect and advise its people while also ensuring that uniformed personnel does not get involved in criminal activity while on duty. The government must play an important role, which brings us to another socio-political context known as bad governance.

### 4.1.3 Misuse of public power

During the most intense October 2020, social media conveyed the immediacy of the protesters' issues in Lagos, Nigeria. The protesters noted that their fight was more than just the #EndSARS; interviewees participants mention that the level of misuse of public power by the SARS operative is the reason why most of them took to the street to protest in Lagos. Misuse of public power failure to bring perpetrators of human rights violators to justice, e.g., SARS violator. Misuse of public power has seemed to put the specific individual above the law they feel untouchable, making some of the SARS operatives carry out various crimes and go unpunished. Misuse of public power violates the right to life; once a SARS officer feels he can get away with anything unpunished, the extrajudicial killings will continue. From the interviews conducted, participant 9, influencer, activist, and online info said that *"Misuse of public power is the single major factor inhibiting police reform. Impunity is fueling outrage and resentment among the populace. Surprisingly, the #EndSARS protest took so long to happen. It should have happened earlier. The systemic use of torture and other ill-treatment by SARS officers for investigations and the continued existence of torture chambers within the Nigerian Police Force points to an absolute disregard for international human rights laws and standards." (participant 9).* A protester affirmed that he was a victim of torture by the SARS unit when given his story I saw the rage and anger he used to express himself to me, according to him *"...their commanding officer (a SARS officer) ordered them to go and hang me. It was then that they carried me to the far end of the hall and tied me up with ropes. Then they began beating me with whatever they could get their hands on, including machetes and rods, afflicting me with all sorts of injuries. One of the SARS men used an exhaust pipe to beat me in the mouth, fracturing one of my teeth and injuring another. I'd been hanging on that hook for more than three hours when I was released..."*

*(*Participant 7*)*; his experience was similar to that of participant 10, who to was a victim of torture *"I am a student studying chemistry in LASU was arrested at home in Lagos, by SARS officers and accused of robbery. I was held in detention for five weeks without access to family, lawyers, or medical care – and was not charged to court. While in SARS detention, I suffered bone fractures and other injuries due to torture and other ill-treatment. I continued assuring them I was innocent. They threatened to shoot me if I didn't confess to taking part in the robbery. I felt weak from not eating since my detention." (participant 10).* I understood from their stories that torture is the intentional inflicting of significant pain or suffering the SARS operative used to put pressure on the victim to make a confession, disclose information, or suffer consequences for their crime. Torture most at times leads to the victim's death, and since most of the supposed criminals have no proper records or identity, they are tagged missing or just forgotten. I also narrated my ordeal with the SARS, which I shared with participant 10 though my issue did not involve torture; in my case, I was arrested at night at about 10 am taken to the station in one of the divisions locked up for a day, I had to bribe one of the officers on duty which he released me without the knowledge of his colleague that proved to me that most of the criminals arrested can be swapped or removed with other victims inside the cell. According to participant 15, *"Misuse of public power sends the message to torturers that they will get away with it. Misuse of public power denies victims and their relatives the right to have the truth established, the right to see justice served, and the right to reparations. No circumstances whatsoever may be invoked as a justification of torture. In many cases, the victims are the poor and vulnerable, easy targets for law enforcement officers whose responsibility is to protect them." (participant 15)*

Observation during the interview with protesters showed the level of intimidation encountered at the hands of the SARS units, which led to them being wrongly extorted by the SARS officer, leading to corruption in another socio-political context.

**4.1.4 Corruption**

Corruption is when a person or group in a position of authority engages in corrupt behavior to get unlawful advantages or misuse power for personal gain. Bribery and embezzlement are two common forms of corruption. Nigeria has a significant issue with SARS corruption at all levels, which affects the whole country. Senior SARS officers loot vast public funds from the same hierarchy level. Meanwhile, at the lowest levels of the SARS hierarchy, rank-and-file officers routinely extort money from the general population, and crime victims are required to pay bribes before the SARS would take their cases seriously. Most of the time, the criminals or arrested individuals are asked to pay a large sum of money for their bail.

Armed SARS officers physically assault ordinary Lagosians working as cab drivers, market traders, and shopkeepers every day, demanding bribes and abusing their human rights to get money. Failure to pay often results in arrest and physical violence. These threats are frequently carried out. Victims of crime must pay the police from the minute they visit a police station until the day their case is taken to court. High-ranking cops steal vast amounts of taxpayer money intended for basic operations. Senior SARS also impose a bizarre system of "returns" that forces rank-and-file cops to give over a percentage of the money they extort from the people. Authorities responsible for policing have been ineffective for years, increasing impunity for officers of all grades who frequently commit crimes against the population they are sworn to protect. For instance, participants 12 and 11 had a similar story of SARS corruption *"I am a taxi driver shuttling day and night for passengers around Lagos. At times, I fall into the hands of SARS operatives who demand I provide my particulars (car papers). When they still check, they still demand I give them something, and on days that I default, I pay the large amount that is more than what I make some days." (participant 12)* while participant 11 narrated his encounter with the SARS operative *"My cousin was a victim of a crime which the SARS operative alongside his friends, I was called by SARS operative in charge of the operation which they told me the situation on the ground on reaching the station, the person in charge obliged I pay the SARS* *officer to release my cousin and avoid taken the case further." (Participant 11).* Aside from this, Participant 14 had a different encounter with the SARS operative regarding corruption. According to him, "*I was traveling from Ibadan to Lagos an encountered a checkpoint, the SARS operatives checked my papers and found out something was missing the beginning of extortion was asked to pay a sum of ten thousand nairas or suffer delay."(Participant 14)* when he was narrating his story, I understood that the SARS units overstepped their boundaries by doing a specific duty meant for the road safety and vehicle inspection officer. According to an activist participant 13, who happens to be a participant, he said that "*SARS checkpoints for motor vehicles, where money is required to avoid harassment and delay, are common sights along major roads in Lagos, Nigeria. Often, they arrest, detain, torture, and kill for a bribe at their "extortion roadblocks" when their demands are not met by the drivers of taxis, minibusses, and motorcycles, as well as private motorists."* (Participant 13).

SARS Corruption erodes public trust in the SARS, degrades legal respect, weakens departmental discipline, and damages SARS morale. The youths protested because most of them from the interviews showed that extortion against their will happened and the only way to voice their grievances social media videos and picture sharing was the only way which then resulted to protest.

### 4.1.5 Unemployment

Unemployment means when someone is actively looking for work but cannot find any, that person is unemployed. Nigeria is one of the most populated nations in Africa. Lagos tends to be the center of excellence, with most youths settling in Lagos seeking greener pasture or means to survive. During the research, it was gathered that the majority of the youths in Lagos, Nigeria, have no paying jobs most depending on the street to survive, these are youths most at times that are graduates either master's holder, degree holders, diploma holders, and secondary school certificate holders. Research has shown that the government of the day gained power from the promises they made during campaign change in 2015, which brought them to power a change the youths felt did not happen. A community's ability to thrive and flourish economically may be harmed if many young people are unemployed. Young people without jobs are more likely to feel excluded from society, leading to worry and a lack of hope for the future if unemployment isn't addressed. For instance, a protester confirmed during an interview, "*An idle man is the devil's workshop, to eat a normal square meal is a problem am tired of not getting a good-paying job for me a reason I joined the protest."* (*Participant 16)* the way and manner the participant said it proved to me that we are seriously seeing a generation of youth that will get at nothing to see their lives back in shape. According to an activist, he said that *"The moment the hoodlums realized the police had been overwhelmed they took over control of the state and unleashed havoc. Most of these hoodlums are unemployed youths crying out for change." (Participant 13)*; meanwhile, participant 17 had a different view from that of participant 16 in his opinion *"In addition to the injustice caused by SARS, a nation that is home to so many unemployed and impoverished people can hardly be called peaceful. After years of bottled-up rage, the lack of effective leadership at a crucial juncture fanned the flames of poverty and unemployment among young people. However, even among employed young people, poverty is still a reality for many." (Participant 17).* Research has shown that the level of unemployment in the country is the cause of so many underlying issues the youths are facing. Unemployment is the cause why most of the youths turned into a crime; some have indulged in cybercrime, also known as yahoo.

The 'Yahoo-Yahoo,' also known as cyber fraud under Nigerian law, is not new. The illegal trade, primarily carried out by young men and women, is a throwback to the '419' work of the 1980s and 1990s. Unemployed youth tend to do anything within their power to make ends meet. According to participant 18, "*People who believe they are financially weak or marginalized, or who believe they are sliding* *through the cracks of society, are more inclined to prioritize their present needs above the implications of their actions in the future."* (participant 18) unemployment in Lagos has been the bedrock of so many crimes committed in most areas; most of the street crimes, according to participant 20, showed how unemployment yielded more criminals. According to him, *"covid-19 has resulted in financial hardship for many people across Lagos while so many lost their jobs because companies had low return other worked without pay so to make ends meet most of the youths resulted to crime." (Participant 20)* some of the unemployed youths during the #EndSARS protest got involved because they were idle and jobless; some of them felt the avenue of the protest could change certain situations the country is facing an interview with a participant 19 who is a job seeker for the past nine years since graduation said that "*I am a graduate of geography from Lagos state university ever since I graduated it has been a struggle, for me now I do anything possible to fend and cloth myself that has turned me to something I sometimes wonder if am the person doing some of the stuff I do."(Participant 19)* I went further to ask him what he does to survive, and he confided in me, telling me that he and most of his friends paid to do stuff for politicians during the election. Unemployment has resulted in the youths losing their societal numb, making them do anything possible to make them survive. With the protests, young Nigerian unemployment has been brought to light, and the need to develop a long-term strategy to harness that energy and use it to help the nation grow, flourish, and advance. The level of unemployment resulted in the anger that developed between the SARS officers. The youth's observation shows that when the SARS operative tries to extort from the unemployed youths, the wrath of giving what you can't afford has resulted in an argument between the youths and the SARS officers, resulting in the officers using force turn result to SARS brutality. Research has also shown that the unemployment level heeded to insecurity.

### 4.1.6 Insecurity

Insecurity by context is a direct effect of dangers and vulnerabilities presented by the possibility of being a victim of crime. Security problems in Nigeria are confronted with several outstanding overlapping issues from extremist insurgencies, nomadic and farmer's confrontations, and kidnapping. Almost every area of the nation has been touched by violence and criminality. For instance, an interview conducted with participant 21 shows that insecurity threatens the society "*with every attack human lives are lost or permanently damaged, and faith in democracy and the country is diminishing*" (*Participant 21)* when the present administration came into power in 2015, pledges were made to safeguard both lives and properties of its residents from terrorist and criminals, yet less than two years before the end of their reign the nation is more unstable than the previous decades. Insecurity in the nation has been related to rising poverty, which is driven by rising unemployment, and the country's economic decline, which has been blamed on the covid-19 pandemic. According to participant 22, he believes that the reason the #EndSARS took another dimension was as a result of *"the youth have their grievance; they have their reasons for going into crime. I do not advocate for anybody to go into crime. But hunger breeds anger, and anger, breeds bitterness. So, I see young men and women who are bitter with the system." (Participant 22)* Various types of instability have emerged, ranging from the Boko haram, who has tormented the country's northern region, to the herders and farmers at odds over land, water, and grazing routes for the herders' animals, among other things. According to studies, most nomadic animals in the Sahara Desert migrate south or west due to climate change, and as a result, confrontations always occur or erupt. When such clashes occur, lives and property are lost, leaving the nomadic farmer without a means of subsistence, which results in kidnapping and extortion. Because they are familiar with the terrain, it is always easy to abduct individuals and hide them, most often in hidden hideouts within the bushes. Banditry and kidnapping are now among the scariest threats for families in Lagos, Nigeria. For instance, participant 23 said that *"My* *mom has developed high blood pressure, she calls me almost every day, weeping because of the nature of my work I close late and coming back home after work is scary for her. Insecurity has changed her perception about my job." (Participant 23)* school children from their classrooms and boarding houses are abducted and released after a ransom is paid. When the kidnapped persons' relatives fail to comply, the abducted is killed, and his parts are sold to ritualists and individuals who buy human parts. Kidnapping has proven to be a lucrative business, and it's now expanding beyond the control of the government. Insecurity has now posed a real threat to trade, education as well as farming, according to participant 24, who said that *"Schools shutting down and calendar changed because of the level of kidnapping going on in the country, we are living in a country where almost every working."* (*Participant 24)* the country has experienced so many setbacks which protesters through #EndSARS with the help of the social media platforms have drawn the attention of the general public both international and local. Attack by bandits has resulted in many fleeing their homes for safety. For instance, participant 25, who is a victim of insecurity, narrated the story of how her uncle was kidnapped and that led them to sell properties to secure his release "*my uncle was kidnapped, and a lot of money was demanded of us, we had to sell some his cars to pay for the ransom, that has become the country we find ourselves today. In my opinion, if the government can't guaranty our safety, then what can we the civilians do." (Participant 25)* during the interview session, I observed that she has already given up hope in the way she complained about the recent happenings of the country.

**4.2 Everyday Life Experiences on Social Media Platforms and in Real-Time**

The second factor identified during the study is everyday life experiences, i.e., experiences on social media platforms in real-time social contexts. Many people's life now revolves around social media. The study's found that half of the youngsters aged eight to seventeen had social media accounts like Facebook, Instagram, TikTok, Twitter, Snapchat, etc. People of many ages and professions use social networking platforms to communicate. Social media has become a vital aspect of any platform in recent times. Social media dominates our daily lives. When one posts on social media, it dramatically influences individuals. Staying in contact with loved ones is essential in our lives, and social media has helped us achieve that. These are only a few of the many factors that raise people's knowledge of social media in general. It not only allows you to stay in touch with friends and family, but it also provides you with a lot of helpful information. Global news sources are at your fingertips with a single mouse click. Everything must be available online to stay up with the rapid changes in technology. Social networking has emerged as a critical marketing tool for web-based enterprises. Newspapers are no longer used as a source of news or advertisements. An alternative is to use well-known social media platforms like Twitter, Facebook, and Instagram to show all of your ads.

While EndSARS protesters experienced the socio-political context mentioned above physically, others learned about it by observing what was happening to the people they came in contact with or as they went about their daily activities while using their digital devices to gain access to uploads and complaints by individuals, the everyday life experience of citizens of Lagos, Nigeria through globalization during the protest shows how events unfold in real-time as they occur. Fundamental changes have occurred across society due to information technology, propelling it from the industrial to the networked eras. Global information networks are critical infrastructure in our world—but how has this altered human relations? The Internet has transformed business, education, government, healthcare, politics, and even our relationships with our loved ones—it has become one of the primary drivers of societal development.

Technology has changed the perception of life; people use technology as if it has been infused into their usual way of life. Information on EndSARS had been circulating through technology without humans knowing we were already preaching EndSARS. For example, individuals experienced daily socio-political contexts (SARS brutality, bad governance, misuse of public power, unemployment, corruption, and insecurity). People were suffering in real-time, and now social media has become a platform to voice their discontent and grievance. As the complaints grew due to more knowledge shared on social media, most people became aware of the events in Lagos, Nigeria. For instance, because social media through Twitter promoted the #EndSARS, the CEO of Twitter, Jack Dorsey, pledged his support by donating a link associated with the feminist coalition, one of the prominent groups supporting protesters.

Social media platforms have fundamentally altered how individuals choose to interact and cooperate. An online community is a socio-technical area in which a feeling of social identification fosters involvement, which increases satisfaction. In situations, social media usage implies that general user reactions vary from those in a crisis, becoming more sensitive. In times of crisis, a shortage of information sources paired with too many user-generated status reports may fuel a rumor mill. Emergency responders must distribute facts while suppressing misinformation threads to regulate collective fear in society. During the EndSARS protest in Lagos, Nigeria, most social media participants online learned about the happenings from ongoing trends shared through social media platforms. For instance, an interview with participant 2 showed that "*I first heard about the SARS brutality on Twitter through the Hashtag #EndSARS; initially I didn't take it seriously until I came across a negative content (video) of how the SARS team brutalized a youth on Twitter" (Participant 2).* The need to communicate and progress in digital technology have spurred the rise of social media. It's about making and keeping intimate relationships at scale. While mobile phones and tablets were first used for desktop or laptop-like experiences, the availability of high-speed wireless internet in homes, workplaces, and public locations increased. With mobile social media applications, users can take their networks wherever they go. A few years ago, social networking was created to connect digitally with people they would not have met otherwise. This change in emphasis from desktop to mobile was made possible by the introduction of smartphones. Apps like Facebook, Twitter, Snapchat, Instagram, and TikTok prospered. Video and photos became the center of mobile applications as technology improved, compelling in-phone cameras. Users may now broadcast live messages in addition to textual communications.

### 4.3 Theoretical Elaboration of Research Findings

### 4.3.1 Nigerian Socio-Political Context and the EndSARS Protests

The Nigerian socio-political context adopted by this research is bad governance, which resulted from the government's quality of services. According to George-Genyi (2014), bad governance involves many problems, including implementing unjust policies or deception inside a system, a lack of transparency and accountability, uncontrolled policymaking, and the illusion of governed people. Bad governance is seen in politicians' strange lifestyles. People's lives and property are unprotected, and economic and social opportunities are not available (World bank, 2020). It is challenging to consolidate democracy when the government cannot meet popular demands. Bad governance is like calling a circle a square. For lack of good leadership, residents are poor despite the nation's abundance of human and natural resources. A government that cannot protect democracy cannot assure its consolidation. Bad governance has always been an issue in Nigeria since the military regime. The Nigerian government could have averted the problem of bad government by introducing technology that could have combated bad governance. Such as the transition from government to e-government. To regulate human relationships, interactions, and activities requires governance in all civilizations. Using governance is like navigating a ship in this situation. Seamanship is needed to guide a ship in a specific direction. In this way, governance as a process involves developing individuals (Misuraca, 2007). Thus, e-governance has been recognized as having a positive influence on individuals. e-Government uses information and communication technologies (ICTs) by government and people. The government uses ICTs to encourage individuals to engage in governance and make the government more open and responsible to the general public (Fatale. J, 2012).

Another Nigerian socio-political context adopted is SARS brutality which is the purposeful use of excessive force against civilians by police or security forces (Socyberty, 2010). SARS brutality has led to numerous death of the citizens the police were meant to govern. According to Human Rights Watch (2018), unnecessary and unlawful weapon use is not uncommon among Nigerian police. Illegal killings often occur during police operations. Sometimes police shoot and kill motorists who refuse to pay bribes at checkpoints. It is rare for police to accidentally shoot and kill someone on the street believing they are "armed robbers." Many persons go missing while in police custody, and their deaths are highly suspected. Despite their crimes, no SARS personnel have been convicted since 2017, which is surprising given the abundant proof of torture. SARS has had the freedom to handle inmates and has gotten protection from the government and the courts, even though their activities are illegal under Nigerian law (Amnesty,2017). However, SARS brutality would have been reduced if technology had been introduced to the system. Studies showed New York City's reaction to the July 2014 police-induced murder of Staten Islander Eric Garner. He was stopped for a non-violent crime and ultimately killed by what was perceived as a brutal police force. The event was captured on video and seen over 3 million times on YouTube. Mayor Bill de Blasio called the event an "awful tragedy" that highlighted poor police-community relations (*Government of New York City, 2014). The NYPD released a thorough action plan for interacting with community residents to discourage crime in June of 2015. According to the proposal, body cameras will be used to better monitor police-citizen interactions in the future; using the body camera; SARS brutality will be reduced drastically.*

*Misuse of power is another factor adopted. It is* authority to gain an unfair advantage over people, groups, or governments. Misuse of power can be seen from different angles depending on the context of the study. The misuse of power by the government has to do with the government setting outlaws that they failed to uphold or the uniform personnel who act above the law and still get away with it. According to Karl Marx conflict theorists, exist to maintain social inequality and to restrict further and repress the lower classes, who are estranged from socio-political and economic resources by the upper classes. Studies have shown that the lower class in society is always the victim of oppression and is always at the receiving end when discussing the misuse of power. According to Ihonvbere (2009), the elite controls and dominates the economic commanding heights and defines society's intellectual and philosophical orientation. For a society like Nigeria to be functional, the government should improve its governance and create technological platforms where complaints can be viewed and tackled. Thus, e-governance has been recognized as having a positive influence on individuals. e-Government uses information and communication technologies (ICTs) by government and people. The government uses ICTs to encourage individuals to engage in governance and make the government more open and responsible to the general public (Fatale. J, 2012).

Corruption another factor adopted is when people or organizations who are in positions of authority are guilty of corruption when they dishonestly seek to earn unfair advantages or misuse their work for their profit, which is a type of dishonesty or a crime, bribery, and embezzlement (World bank, 2020). Corruption has hindered the country's development and has given Nigeria a negative impression from the rest of the world. Studies show that Nigeria ranked 149th out of 180 nations in Transparency International's Corruption Perception Index 2020, scoring 25 out of 100. Nigeria is currently the second most corrupt nation in West Africa (Johnson, 2021). For A country like Nigeria to combat corruption, strict measures need to be implemented. Police corruption is the act of extorting or getting paid to ignore a crime or a potential future crime. For example, bribes might protect unlawful activity or sell criminal information. Bribery is a widespread kind of corruption amongst police personnel. A shift in an officer's conduct inside the department's "subculture" may corrupt behavior. A subculture is a group of people who share common attitudes and values. Psychological, sociological, and anthropological perspectives may explain new behavioral evolution among police personnel (Belmonte, 2015). To solve the issue of corruption in the country, ICTs may help combat corruption. New technologies may improve openness, accountability, and civic involvement by enabling information flow between government institutions, citizens, and people. (Chene, 2011). Automation, removing intermediaries, and decreasing red tape are all ways that ICTs may help bring about beneficial change (Zinnbauer, 2012).

Unemployment was adopted, studies show that unemployment is now one of Nigeria's major developmental issues. However, the reflections from different local and international organizations and the clear evidence of unemployment in this decade suggest that there has never been a moment in Nigeria's long history when unemployment is as significant as today. One cannot reasonably say that the Nigerian government has not done anything to eliminate unemployment. Unemployment has resulted in the numerous problems Nigerians are encountering. Unemployment refers to those who are employable yet are unable to obtain work. An employer's failure to fill an open position might lead to a worker's inability to find a job. High-interest rates, a worldwide recession, and a financial crisis may lead to decreased demand. Nowadays, the government relies on the private sector for job creation, while the private sector relies on the government for policies that enable employment creation. But no one has questioned which employment must be generated, for whom, how, or why, particularly to address these inequities. Maybe digital technologies might help us see the future of youth empowerment and employment. First, personal, market and other extensive data collecting must be enhanced, as must data analysis. At this point, the person's employability (soft skills) must be considered. Our society's most significant concerns, such as unemployment or lack of employability, will be addressed as digital technologies grow more efficient and flexible (Duncan, 2014).

Another Nigerian socio-political context adopted is insecurity. Securing is a notion that precedes the state, according to Omoyibo and Akpomera (2013). The state's primary duty is to provide security (Thomas Hobbes, 1996). As stated in Nigeria's Constitution of 1999, "the fundamental goal of government must be to ensure the people's security and wellbeing. Sadly, the government has failed in its constitutional duty to protect lives, property, and economic activity. Anxiety in Nigeria has encouraged crime and terrorist assaults throughout the country, harming the economy and progress. And in 2011, the National Assembly enacted the Anti-Terrorism Act (Ewetan, 2013). Unemployment remains high despite these efforts, as seen by Nigeria's poor rating in the Global Peace Index (GPI) (GPI, 2012). The lack of protection causes worry or anxiety, according to Beland (2005). Insecurity is defined by Achumba et al. (2013) from two angles. On the other hand, the danger is the condition of being vulnerable to damage or injury. Also known as risk or anxiety, insecurity is a condition of being exposed to danger or worry, where anxiety is a vague unpleasant sensation felt in expectation of some catastrophe. These definitions of insecurity emphasize that persons afflicted by it are unsure or uninformed of what will happen and exposed to threats and risks. Insecurity is a breakdown of peace and security that contributes to periodic conflicts and the destruction of lives and property. Nigerian technology expert Muhammad Isma'il has proposed unconventional remedies to Nigerian insecurity. According to Isma'il, Nigeria's insecurity is escalating every day, demanding a comparable technical intervention to developed countries. "According to Isma'il, technology can quickly fix most of Nigeria's security and economic issues. Isma'il says the Nigerian government can employ identity management to "provide social security, peace and order, and a well-organized society."

Following the findings of this study and insights in the extant literature:

*Proposition I: The Nigerian socio-political contexts likely promoted the occurrence of the EndSARS protests.*

### 4.3.2 Nigerian Socio-Political Context and Everyday Life Experiences and the EndSARS Protest

My interrogation of the extant literature reveals that scholars have not used socio-materiality to study protests in Nigeria. The study shows how the Nigerian socio-political context is interwoven with the everyday life experiences of Nigerians in real-time and on social media. It reveals how Nigerians experience the socio-political contexts. The extant literature has provided comprehensive insights into how bad governance has resulted in protests in Nigeria and other contexts. According to George-Genyi (2014), bad governance involves many problems, including implementing unjust policies or deception inside a system, a lack of transparency and accountability, uncontrolled policymaking, and the illusion of governed people. Bad governance also occurs when the people are not given a voice and not cared for. While (UNESCAP, 2011) is defined as the process of making choices and implementing or not implementing those decisions. There are many studies of bad governance in Nigeria in the extant literature. All this literature has shown that Nigeria has been irresponsible because they don't promote policies that have the interest of its citizens at heart. Youths headed by Joe Okei-Odumakin, president of Campaign for Democracy, recently protested in Lagos. The protest resembled one held in Abuja on March 16, 2011, by the Enough is Enough Coalition. Their demands included proper representation of the people in politics, a halt to bad governance, restructuring the nation's security structure, and enough electrical supply. Their needs were clamoring for excellent control from the people that govern. While (Rotberg 2004) tried to compare the relationship between poverty, bad governance, and state failure, this combination creates an ideal breeding environment for militancy and eventual instability. Given that so many researchers are critical of inadequate governance and often perceive it as colonialism, much of what was observed focused on how one group or authority may dominate other people or regions, generally via colonization and economic superiority (Rodney, 2018). Consequently, none has been done to show how socio-materiality impacted bad governance. Most of the studies drew their attention to looting and focused on the government embezzling funds in a real-life social context. However, this study combines the analysis of bad governance from the perspective Nigerians experience in real-time and on social media showing how social materiality can impact the investigation. According to Socio-materiality, humans and technology are intimately connected (Orlikowski 2007; Orlikowski and Scott 2008). (Leonardi et al., 2012; Scott and Orlikowski, 2012).

Another strand of the extant literature related to the findings of this study is Police brutality. This is a factor that leads to EndSARS. The word "police brutality" was first used in the press in 1872. The first occurred at the Harrison Street Police State in Chicago when a citizen was beaten into a coma (Chicago Tribune News, 1999). Brutality is one of the various wrongdoing that includes security officials abusing individuals. This crime known as police brutality occurs in many nations and territories worldwide, particularly in developing ones. Police brutality is the purposeful use of excessive force against civilians by police or security forces (Socyberty, 2010). For example, police brutality might entail physical or verbal abuse, injuries, or death. The special anti-robbery squad (SARS) was formed in 1992 to combat armed robbery and other serious crimes. As used in this context, it is as a result that it was created from a unit of the police force. Scholars who study SARS brutality have shown that SARS, the Special Anti-Robbery Squad, is active throughout Nigeria's southwest, particularly in Lagos. The study has shown that Nigerian citizens are mishandled, beaten, shot indiscriminately, mutilated, and killed in a variety of ways, including; unnecessary restraints like handcuffs and leg chains, unnecessary use of firearms against suspects and innocent members of the public, torture of suspects to forcefully extract information and extortion of satisfaction. (Human Rights Watch, 2018). Unnecessary and unlawful weapon use is not uncommon among Nigerian police. Illegal killings often occur during police operations. Sometimes police shoot and kill motorists who refuse to pay bribes at checkpoints. It is rare for police to accidentally shoot and kill someone on the street believing they are "armed robbers." Many persons go missing while in police custody, and their deaths are highly suspected (Human

Rights Watch, 2018). I observed that most studies on SARS brutality used failed state theory to examine SARS brutality. The term "failed state" typically refers to a state that has failed in some of its fundamental tasks as a sovereign government. A failing state has to crumble social and political systems (Anyanwu, 2005). At the same time, my study is trying to link technology, everyday life experience in real-time, and society. Consequently, social media platforms allow for the integration and replication of resources such as knowledge, skills, and expertise by actors both within and outside the society (Eaton et al. 2015; Gawer 2014; Tilson et al. 2010b).

Misuse of power can be seen as misusing one's authority to gain an unfair advantage over people, groups, or governments. The roots of Nigerian police may be traced back to the colonial era. Colonial policies overthrew the old informal law enforcement system and imposed a western model of police on the indigenous peoples of the region. The police, according to conflict theorists, exist to maintain social inequality and further restrict and repress the lower classes, who are estranged from socio-political and economic resources by the upper classes. Others have written on the cops from this angle. For example, Bowden (1978) argues that the police suppress the poor and weak to preserve the rulers' interests, whereas (Brogden, 1982) says that the police are structured, organizationally, and ideologically to operate against the marginal levels'. Hence, the previous school of thinking argues the police–community connection is not reciprocal. Consequently, western orientation has been linked to policing in the context of oppression and gross misuse of power. There are severe consequences for Nigerian police if this colonial ideology is continued and fostered by future Nigerian administrations. (Ahire, 1991) studies have also shown that the use of excessive force by police against people is not just immoral but also unlawful. In truth, most police-citizen encounters are limited to law enforcement settings. Citizens dislike conditions they see as restricting. (Alemika & Chukwuma, 2000) had a different view of the misuse of power. In their article, they believed that the ineffectiveness of Nigerian police due to lack of human resources, infrastructure, and finances degrades public perception of police. These settings polarize police and citizens, leading to hatred and bloodshed. Most of the study findings used a sociology point of view as a medium to discuss the misuse of power by the police. In contrast, this study uses socio- materiality, which is the intersection of technology, everyday life in real-time, and the society that the police are meant to govern.

Corruption is a subject of interest to a group of researchers engaged in the study of society. According to critics, corruption in Nigeria is a persistent problem, with public officials and others with whom they interact violating established rules and contracts. Poverty, political instability, and other socioeconomic pressures pressure public workers to be corrupt. This is particularly true when officials are aware that their possibilities may be limited as a result of a coup or a loss at the elections, or when their relatives impose more demands on them, or when they feel forced to maintain a detectable level of living (Colins, 1965; Nye 1961). According to Ruzindana (1999), he identified several instances of corruption in Nigeria, including bribery, private gain, and benefits for non-existent employees and pensioners (called ghost workers). Studies have shown that citizens in Nigeria, particularly in Lagos, have lost trust and view the government agent (SARS) as a unit with no integrity. According to Adegoke (2020) when people have a negative view of the police, they are more likely to participate in anti-government rallies in Nigeria. In the absence of high-profile complaints of police violence, poor police service and bribe demands create daily victims of individuals of all socioeconomic backgrounds, particularly the poor.

Unemployment is now one of Nigeria's major developmental issues. However, the studies from different local and international organizations and the clear evidence of unemployment in this decade suggest that there has never been a moment in Nigeria's long history when unemployment is as significant as today. One cannot reasonably say that the Nigerian government has not done anything to eliminate unemployment. Unemployment is now one of Nigeria's major developmental issues. However, the studies from different local and international organizations and the clear evidence of unemployment in this decade suggest that there has never been a moment in Nigeria's long history when unemployment is as significant as now. One cannot reasonably say that the Nigerian government has not done anything to eliminate unemployment. (Aganga, 2010 and Ogunmade, 2013).

According to the international labor organization (ILO), unemployment is among the biggest threats to social stability in Nigeria and other countries. An increasing tendency in unemployment has been one of the significant economic concerns confronting Nigeria. It includes students, homemakers, retirees, and anyone who doesn't want to work or is too old to work. Unemployed persons who cannot work yet cannot find adequate paid labor (Njoku and Okezie, 2011). Unemployment causes widespread poverty, young unrest, high rates of social vices, and criminal activity, and if not managed, indifference, skepticism, and revolution may result. Desperate times bring out the worst in people. Most young graduates arrested in crimes like armed robbery, kidnapping, and prostitution blame their actions on the unemployment crisis in Nigeria. It has been suggested that the high rate of communal crisis and youth restiveness, hire killings and assassinations, kidnappings, and other forms of criminality such as 419 and even the Niger Delta uprising and recent 'Boko Haram' insurgents in the northern part of the country, are due to high unemployment (Kayode, A. et al., 2014). Unemployment is attributed to causes such as rapid population increase, rural-to-urban migration, and a lack of marketable skills and experience among the youths (Ongbali S. O et al., 2019). Based on my observation of the unemployment rate, both studies proved that unemployment is the bedrock of most crises occurring in the country. Nowadays, the government relies on the private sector for job creation, while the private sector relies on the government for policies that enable employment creation. But no one has questioned which employment must be generated, for whom, how, or why, particularly to address these inequities. Maybe digital technologies might help us see the future of youth empowerment and employment. First, personal, market and other extensive data collecting must be enhanced, as must data analysis. At this point, the person's employability (soft skills) must be considered. Our society's most significant concerns, such as unemployment or lack of employability, will be addressed as digital technologies grow more efficient and flexible (Duncan, 2014)

Insecurity is a direct effect of dangers and vulnerabilities presented by the possibility of being a victim of crime. Social, economic, political, and cultural growth can only occur in a peaceful environment. The development would be impossible without a conducive production, industry, and commerce as money for development is diverted to security issues. And private investors will not invest if their money is not safe. The degree of insecurity in Nigeria is worrisome, having infiltrated nearly every element of our national life. (Shettima, 2012). According to Ewetan & Urhie (2014), insecurity started well before the colonial period. Because most of the country's problems stem from that period. Things didn't go as planned throughout that time frame, and grudges built up. Following the Nigerian civil war, vast amounts of weapons were imported for military purposes, some of which ended up in the hands of civilians. Securing is a notion that precedes the state, according to Omoyibo and Akpomera (2013). The state's primary duty is to provide security (Thomas Hobbes, 1996). As stated in Nigeria's Constitution of 1999, "the fundamental goal of government must be to ensure the people's security and wellbeing. Sadly, the government has failed in its constitutional duty to protect lives, property, and economic activity. Anxiety in Nigeria has encouraged crime and terrorist assaults throughout the country, harming the economy and progress. And in 2011, the National Assembly enacted the Anti-Terrorism Act (Ewetan, 2013). Unemployment remains high despite these efforts, as seen by Nigeria's poor rating in the Global Peace Index (GPI) (GPI, 2012). Nigerian technology expert Muhammad Isma'il has proposed unconventional remedies to Nigerian insecurity. According to Isma'il, Nigeria's insecurity is escalating every day, demanding a comparable technical intervention to developed countries. "According to Isma'il, technology can quickly fix most of Nigeria's security and economic issues. Isma'il says the Nigerian government can employ identity management to "provide social security, peace and order, and a well-organized society."

The socio-materiality theory used in this context is the intersection between technology, everyday life experience, and society (Orlikowski, 2007). Socio-materiality has gained popularity among information systems researchers. As a result, stakeholders better know how human behavior is shaped, altered, and organized in the real world. Simultaneously, it allows for detailed technical inquiry and comprehension without certainty (Orlikowski and Scott, 2008). A current trend in information and knowledge distribution, media has elevated communication above conventional communication and socialization constraints, impacting people's social, political, and economic activities. While the Internet was formerly considered a news source, today's civilizations rely on social media for information. Many individuals use social networks to communicate and share material such as text, voice, photos, or videos (Alquraan et al., 2017). They may be utilized for protest, job seeking, socialization, education, entertainment, governance, and political engagement due to their interactive character. Thus, social media as social communication tools encourage involvement, connection, cross-border information sharing, and the development of relationships and interactions between individuals. Facebook, WhatsApp, Twitter, Instagram, YouTube, and Telegram are all popular social media platforms.

The first generation of materialism regarded technology as shaping societal systems. Early e-government studies, for example, had this viewpoint, leading to unrealistic and naive assumptions about technology's influence on government (Bannister and Grönlund 2017). The second generation focused on how people utilized technology and how technology may generate unexpected effects. The technology itself created the consequences, not by its usage (Orlikowsk,2000). According to Misa, determinism and materialism are resolved by prioritizing one level of thought over another. These scholars are "forced to forgo comment on the fascinating topic of whether technology has any impact on anything" because they are "forced to exclude micro-level data" (p. 138). But, as Misa points out, "it seems particularly problematic in an era of global socio-technical problems" to declare that technology has no socially relevant tangible influence on society's orientation (p. 138). Sociologists risk underestimating the importance of technology in history-making. On the other hand, techno-determinists risk imputing motivations and intents without proof. Therefore, humans are reduced to cultural and social victims in determinist worldviews (materialistic or idealistic). Socio-materiality explains how human bodies, social media, and actions are entangled with societal realities. Thes are Nigerian socio-political context is a significant factor identified in the study as the cause of the EndSARS protest.

Following the findings of this study and insights in the extant literature:

*Proposition II: The ways protesters' everyday life experiences were influenced by Nigeria's socio-political contexts and their experiences on social media and in real-time is likely to have promoted the EndSARS protest.*

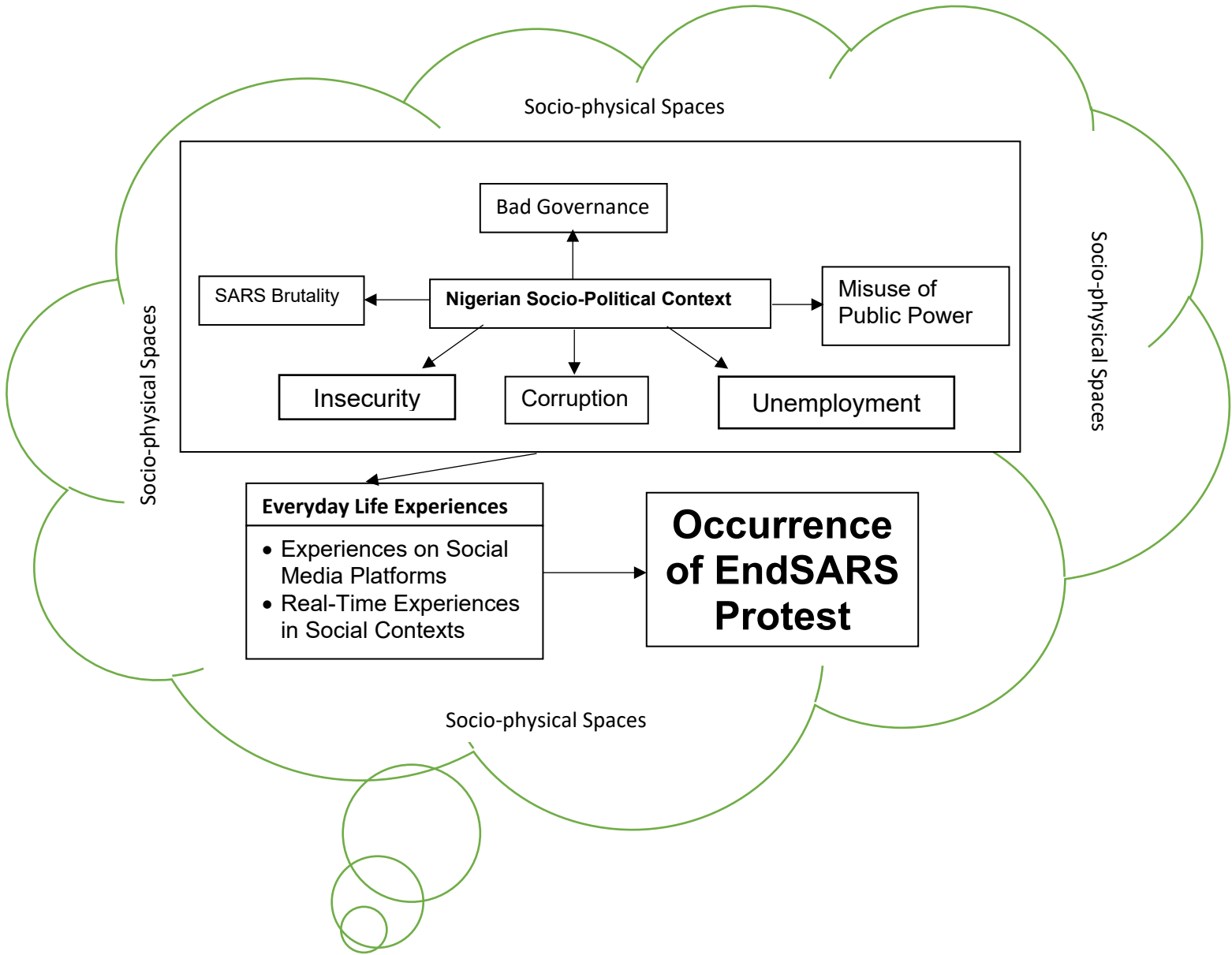

Figure 4.1: Dynamics of how Social Media Impacted the EndSARS Protests

## 4.4 Implication of The Research Model

At the inception of this study, my objective was to investigate how social media impacted the EndSARS protests in Lagos, Nigeria. The research question was: How did social media impact the EndSARS protests in Lagos, Nigeria? Interpretive inductive research approach and snowball sampling technique were implemented to participate in the research process, and the sample populations are directly relevant to the study's objective. The study confirms the strength of the interpretive inductive research approach and snowball sampling technique in gathering different understandings during a protest. Critical issues on the role of social media on EndSARS, such as the Nigerian socio-political context which include the following: bad governance, SARS brutality, misuse of power, corruption, unemployment, and insecurity. This study presents an entirely new perspective on how social media occurred through the following factors. Proposition I shows that those outlining factors of Nigeria's socio-political context lead to the likely occurrence of the EndSARS protest. Proposition II shows everyday life experiences of the people their experience with social media platforms such as Twitter, face, etc. that in real-time as it still can lead to the occurrence of EndSARS,

## Chapter 5
## Summary, Conclusion, and Recommendations

### 5.1 Summary

This study focuses on the role of social media during the EndSARS protest in Lagos, Nigerian, Factors that can lead to the occurrence of EndSARS, and how technology is embedded in human existence in real-time. Because of technological advancements, the Internet, broadband, and World Wide Web have changed the media industry with digital platforms, new processes, methods of communication, and interaction, and in turn, the new media have boosted business opportunities, governance, education, sport, social movements and much more (Idowu and Esere, 2013) Internet and mobile communication technologies played a crucial role in the case of EndSARS. These technologies help open new spaces and modes of social interaction that include decentralized structures, online and offline participation of general people in the case of the related topic. Here, socio-materiality, the intersection between technology, the everyday life of the EndSARS protesters, and the society in which the protest occurred, is being investigated. These technologies have allowed people to organize and challenge the status quo. This study used Socio-materiality theory to examine and evaluate how technology is being used in an individual's everyday life in society. Based on interviews with protesters in Nigeria, this research has helped advance the conversation on how social media might influence government policy and agitate for the end of a unit in police known as SARS in Nigeria. The #EndSARS protest by Nigerian youth may undoubtedly promote youth-led social media peaceful protests throughout Nigeria, which can establish constructive policy processes and good governance on the African continent. The research results show that Nigerians can utilize social media platforms to effect desired changes in their socio-political development. Aside from that, the research has revealed how ordinary citizens in Lagos states may peacefully organize themselves to exert pressure and support on their government to improve governance.

## 5.2 CONCLUSION

An expression of displeasure about a topic or occurrence is called a protest. Strikes, hunger strikes, public rallies, such as the EndSARS in Lagos, Nigeria, and other forms of protest are all visible. As a result of dissatisfaction, individuals address a problem in various ways. In this instance, Nigerian youth demonstrated their unhappiness when killed and demanded justice via protest. So the analysis concluded that most demonstrators were young people. Furthermore, the study found that the #EndSARS protest in Nigeria was sparked by several causes, including faulty government policies, SARS brutality, misuse of power, corruption, unemployment, and insecurity. The study also showed how everyday life experiences of those factors that led to the protest triggered the youths to action. The research found that persons who participated in protests in Nigeria were informed about the events through social media platforms. The study also found that social media platforms are ideal mediums where individuals protest for important causes. The research outcomes showed that social media is an interactive medium where people address their concerns. It is concluded from the research that people in Nigeria would continue to exert pressure on the government to solve public problems if the government that is supposed to rule does not put in place appropriate steps for dealing with society's influencing variables.".

## 5.3 RECOMMENDATIONS

Based on the findings, the following recommendations are put forward:

1. The dualist notion is a recommendation to scholars, scholars should understand that there is no dualism everyday life experience of people and technology are not separate. By promoting the dualist notions scholars are misleading stakeholders including government to indicate that people everyday life experience are different from social media.

2. Technology determinism address government notion on the role of social media in protest. Government by adopting technology determinism are assuming that social media is the cause of protest. Whereas this study has shown that it's not social media that's causing the protest but the socio-political context. Government should know that even if social media is banned protest will still occur because what causes the protest is the socio-political context of Nigeria.

3. Public officials may use social media to gather public opinion. Thus they should support free speech and allow these perspectives to drive their work.

4. For successful participatory government, civil society organizations in the nation must enlighten their members on the use of social media

5. It is essential to perform additional studies in this field because of the ever-changing nature of social media and how it influences the dynamics of society. Social media is also an essential factor in studying mobilization and socio-political concerns.

# Appendix A – Interview Guide

Interview Guide These interview questions aim to investigate how social media influenced the EndSARS protest in Lagos, Nigeria. The questions have been structured to identify your opinions on social media and technology and what you think the protest is all about.

**Introduction**

- **Understanding**

- Brief about self (interviewee)
- Aim of visit
- Statement of problem: Purpose of interest (interviewer)
- Type of interview:
- Time duration (intended):
- Ethics (Skip uncomfortable questions)
- Any question (Is there anything you would like to ask before we begin)?

1. **Getting to know the participant**

- Brief about self (interviewee)

(Name, age, marital status, other interest).

- Do you have stable access to social media and / the internet?
- Q. How often do you use your digital device and/ or the internet?
- If you own a smartphone or any other digital device what do you use it for most of the time?
- Q. when connected to the internet, what app do you often go to for information?
- Q. When did you first hear about the #EndSARS protest, and which app or platforms do you use?

**MAIN INTERVIEW CONTENT**

**Personal**

1. **Role of social media and how it impacted street protest during the EndSARS protest in Lagos, Nigeria.**

- Computer/phone usage in online Protest
- Do you use a computer or/and your phone during the EndSARS protest?

(like chatting, video sharing, or getting information, e.g., NEWS)

(Care to share how you use your phone/computer for socializing?

- On a personal level
- Do you use a computer or/and your phone to follow up on the EndSARS protest?

(like chatting, video sharing, or getting information e.g. NEWS)

(Care to share what you use your phone/computer for as regards to last year's EndSARS Protest?

v If not: Q. may I ask why?

- If using any:

1. what started the #EndSARS protest?
2. what is your opinion that triggered the EndSARS protest?
3. What do you think the #EndSARS protests were about?
4. Can you describe your role during the EndSARS protest?

- **Social media role played:**

**Q.** Do you think social media heated the protest more?

**Q**. According to your judgment which social media platform have a higher impact on the protest

**Q**. What role did social media play during the protest in your opinion?

**Q.** Is there any form of bullying encountered online during the protest, and how did you handle it?

**Q.** Do you know anyone involved in the protest that was bullied directly or indirectly?

**Q.** How do you think the cyberbullying effect can be drastically reduced?

1. **The use of technology during an online street protest during the EndSARS protest in Lagos, Nigeria.**

- Q. What is your opinion about using technology like mobile phones or digital devices during online protests instead of normal gathering and activating an offline protest?
- If a protest were organized today, where is the first place you would hear about it?
- If technology was ever used
- Q. How likely are you to participate in an online protest?

If not: Q. may I ask why?

- Q. How likely would you participate in an offline protest?

If not: Q. may I ask why?

- Do you think that tweeting or sharing public information about a protest can impact fixing societal and political issues?

1. How did this event of the EndSARS Protest change our social and political issues?

- Q. Do you sometimes feel that the Nigerian government tries to silence our political opinions online?

If not: Q. may I ask why?

- Do you know exactly what happened during the #EndSARS protest on the 20th of October 2020 at the Lekki toll gate?
- If Yes or if you have a bit of an idea, do you think there is some false information about what happened on that day?
- Following the Lekki toll last year, would you come out to protest if one is organized today?

If not: Q. may I ask why?

1. **Outro.**

- Q. Is there anything you think I avoided during the cause of my interview that you wish to draw my thoughts regarding the EndSARS protest in Lagos, Nigeria?
- Q. Any questions?  Gate event

THANKS

## Appendix B - Ethical Certificates Human Research

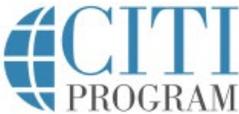

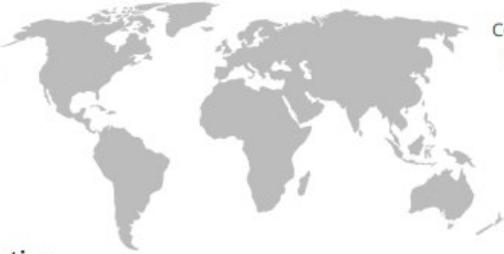

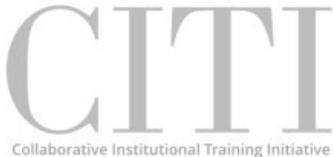

Completion Date 07-Feb-2021
Expiration Date 07-Feb-2023
Record ID 40782124

This is to certify that:

**Christopher Augustine**

Has completed the following CITI Program course:

Not valid for renewal of certification through CME.

**Human Research**
(Curriculum Group)
**Group 2 - Investigators**
(Course Learner Group)
**1 - Basic Course**
(Stage)

Under requirements set by:

**West African Bioethics Training Program**

Collaborative Institutional Training Initiative

Verify at www.citiprogram.org/verify/?w02859805-6961-4628-9ab1-198c2b642d1b-40782124

## Appendix C - Permission to Carry Off-Campus Research

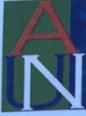

# Appendix D - Faculty Advisor Assurance

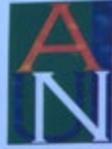

American University of Nigeria – Institutional Review Board
**Human Subjects Protection Program**

**Faculty Advisor Assurance**

Faculty advisors (supervisors of undergraduate or graduate student research projects) are required to read and DIGITALLY sign this statement in order for their students to be permitted to conduct research. *Students, be sure to submit this page, signed, via E-mail along with your entire application to the IRB.*

I am the faculty advisor/supervisor for the student submitting the attached research application. By my signature, I certify that I have reviewed the research protocol and believe that it is scientifically and ethically sound. Furthermore, I believe that the student has the necessary training, experience, and knowledge to conduct the research in a manner consistent with the regulations governing human-subjects research and sound research principles.

*I agree to:*

- Oversee and monitor the conduct of this research by communicating regularly with the student investigator;

- Assist with the resolution of any problems or concerns encountered during the research; and

- Assure that the AUN IRB is notified in the event of a protocol deviation or an unanticipated problem involving risks to subjects or others.

I understand that as faculty advisor, I am responsible for the conduct of this research.

CHRISTOPHER AUGUSTINE

**Student Name**

INVESTIGATION OF HOW SOCIAL MEDIA INFLUENCED THE ENDSARS PROTEST IN LAGOS, NIGERIA

**Title of Research Project** *(should correspond to the title on the application form)*

Dr. Samuel Utulu

**Faculty Advisor Signature**

16/11/2024

**Date**